\documentclass[twocolumn,english,aps,prd]{revtex4-1}
\usepackage{CJK}
\usepackage[T1]{fontenc}
\usepackage[left=1.5cm,top=2cm,right=1.5cm,bottom=2cm]{geometry} 
\usepackage{array}
\usepackage{textcomp}
\usepackage{multirow}
\usepackage{amsmath}
\usepackage{graphicx}
\usepackage{hhline}
\usepackage{comment}
\usepackage{adjustbox}

\usepackage[utf8]{inputenc} 

\usepackage{amsmath}
\usepackage{amsfonts}
\usepackage{amssymb}
\usepackage{cancel} 
\usepackage{youngtab}
\usepackage{xfrac}
\usepackage{subfigure}

\usepackage{graphicx}
\usepackage{feynmp} 
\usepackage{subfigure} 
\usepackage{float} 

\usepackage{pgfplots}
\usepackage{xparse}

\usetikzlibrary{positioning,arrows,patterns}
\usetikzlibrary{decorations.markings}
\usetikzlibrary{calc}

\tikzset{
  photon/.style={decorate, decoration={snake}, draw=black},
  fermion/.style={draw=black, postaction={decorate},decoration={markings,mark=at position .55 with {\arrow{>}}}},
  vertex/.style={draw,shape=circle,fill=black,minimum size=3pt,inner sep=0pt},
  vacuum/.style={draw,shape=rectangle,fill=black,minimum size=3pt,inner sep=0pt},
}

\NewDocumentCommand\semiloop{O{black}mmmO{}O{above}}
{%
\draw[#1] let \p1 = ($(#3)-(#2)$) in (#3) arc (#4:({#4+180}):({0.5*veclen(\x1,\y1)})node[midway, #6] {#5};)
}

\makeatletter

\@ifundefined{date}{}{\date{}}

\usepackage{float}
\usepackage{textcomp}

\usepackage{babel}

\IfFileExists{lmodern.sty}{\usepackage{lmodern}}{}

\usepackage{babel}

\@ifundefined{showcaptionsetup}{}{%
 \PassOptionsToPackage{caption=false}{subfig}}
\usepackage{subfig}
\makeatother

\begin{document}

\title{A $\mathrm{U(1)}$ non-universal anomaly-free model with three Higgs doublets and one singlet scalar field}

\author{S.F. Mantilla}
\email{sfmantillas@unal.edu.co}
\author{R. Martinez}
\email{remartinezm@unal.edu.co}

\affiliation{\textit{Departamento de F\'{i}sica, Universidad Nacional de Colombia,
Ciudad Universitaria, K. 45 No. 26-85, Bogot\'{a} D.C., Colombia} }
\date{\today}

\begin{abstract}
The flavor problem, neutrino physics and the fermion mass hierarchy are important motivations to extend the Standard Model into the TeV scale. A new family non-universal extension is presented with three Higgs doublets, one Higgs singlet and one scalar dark matter candidate. Exotic fermions are included in order to cancel chiral anomalies and to allow family non-universal $\mathrm{U(1)}_{X}$ charges. By implementing an additional $\mathbb{Z}_{2}$ symmetry the Yukawa coupling terms are suited in such a way that the fermion mass hierarchy is obtained without fine-tuning. The neutrino sector include Majorana fermions to implement inverse see-saw mechanism. The effective mass matrix for SM neutrinos is fitted to current neutrino oscillation data to check the consistency of the model with experimental evidence, obtaining that the normal-ordering scheme is preferred over the inverse ones and the values of the neutrino Yukawa coupling constants are shown. Finally, the $h\rightarrow \tau\mu$ lepton-flavor-violation process is addressed with the rotation matrices of the CP-even scalars, left- and right-handed charged leptons, yielding definite regions where the model is consistent with CMS reports of $\mathrm{BR}(h\rightarrow \tau\mu)$.

Keywords: Flavor Problem, Neutrino Physics, Extended Scalar Sectors, Beyond Standard Model, Fermion masses, Inverse See-Saw Mechanism, LFV.
\end{abstract}
\maketitle

\section{Introduction}
Although the Standard Model (SM) \cite{glashow1961partial,salam1968elementary,weinberg1967model} (SM) has been successful to explain the experimental low energy observations in particle physics, there are some theoretical and observational evidences that suggest an underlying Beyond Standard Model (BSM) extension. Two of these evidences are the fermion mass hierarchy problem and the neutrino oscillation. In the hierarchy problem \cite{georgi1986flavor}, the mass of the fermions and their mixing requires arbitrary fine-tuning of the Yukawa coupling constants. Some approaches in the framework of BSM extensions involves schemes to explain this puzzle in the framework of zero-texture structures of the Yukawa matrices \cite{fritzsch1978weak, fukugita1993phenomenological}. Moreover, in these schemes, the neutrino oscillation problem could be addressed, obtaining satisfactory models of flavor physics. 

The confirmation of neutrino oscillations and the massive nature of neutrinos have been widely confirmed by precision measurements done by a huge number of experiments \cite{davis1968search,abdurashitov2009measurement,kaether2010reanalysis,cleveland1998measurement,aharmim2013combined,
bellini2010measurement,borexino2014neutrinos,hosaka2006solar,hosaka2006three,cravens2008solar,aartsen2015determining,
gando2011constraints,apollonio1999limits,piepke2002final,an2015new,kim2016measurement,kopp2013sterile,adamson2013measurement,
adamson2013electron,abe2014observation,kolupaeva2016current,adamson2017constraints} as well as their mixing angles. The references \cite{esteban2016updated,nufit} show the most recent fit from the experimental data. The massive nature of neutrinos motivates different scenarios BSM where the origin of the smallness of their masses could be understood. The preferred mechanism to obtain small masses is the \textit{see-saw mechanism} (SSM) \cite{minkowski1977mu,gell1979r,yanagida1980horizontal,mohapatra1980neutrino,schechter1980neutrino} which introduces new Majorana fermions with their corresponding mass terms in the Lagrangian in such a way that the enormous scale of their masses ($10^{14}\,\mathrm{GeV}$) suppress the electroweak ones, yielding small neutrino masses of the active neutrinos at the eV scale. However, the large scale of the Majorana neutrinos is unreachable by current or future high energy experiments. There exists another mechanism, the \textit{inverse SSM} \cite{mohapatra1986mechanism,mohapatra1986neutrino,catano2012neutrino,dias2012simple} which introduces additional neutrinos that reduces the Majorana mass scale into the experimentally accessible energies. Also, a Majorana neutrino could induce matter-antimatter asymmetries through the \textit{leptogenesis mechanism} \cite{fukugita1986barygenesis}.

On the other hand, there are different models BSM with extended scalar sectors. The interest by these type of extensions has grown after the detection of the Higgs boson at ATLAS \cite{aad2012observation} and CMS \cite{chatrchyan2012observation} experiments at the Large Hadron Collider (LHC), being the best known the two-Higgs-doublet model (2HDM) which introduces two charged $H^{\pm}$, one CP-odd $A$ and two CP-even $h$ and $H$ scalar bosons by proposing the existence of a second Higgs doublet \cite{gunion1990higgs}. Such models arises naturally in supersymmetric (SUSY) extensions. Also, 2HDM yield scenarios where the large hierarchy between the $t$ and $b$ quarks can be understood by proposing a vacuum hierarchy between the two doublets \cite{branco2012theory}. Other models extend to the Next-to-Minimal 2HDM (N2HDM) by adding to the minimal 2HDM a scalar SM-singlet which could yield the spontaneous symmetry breakdown (SSB) of additional $\mathrm{U}(1)$ gauge symmetries. Another scenario proposes an additional scalar field as candidate to be dark matter (DM) \cite{he2010simplest,grzadkowski2010tempered,profumo2014fundamental,he2016new,han2017higgs,
kumar2016phenomenology,carena2016alignment} which does not receive vacuum expectation value (VEV).

Regarding the abelian extensions of the SM \cite{langacker2009physics}, different issues can be addressed as neutrino  physics \cite{ma1996neutrino,barger2003primordial,king2006theory}, flavor physics \cite{langacker2000flavor,leroux2002flavour,baek2006b_s} and DM phenomenology \cite{hur2008supersymmetric,belanger2008dirac,profumo2014constraining,alves2015dirac,alves2016r,arcadi2017waning}. The direct consequence to add an abelian gauge symmetry is the appearance of an additional neutral gauge boson $Z_{\mu}'$ which may modify some electroweak observables \cite{leike1999phenomenology,accomando2016z,rose2017explanation} through the mixing with the ordinary $Z_{\mu}$ boson after two SSBs, the first one triggered by some scalar singlet and the second one being the electroweak SSB. 
Other extensions are non-universal of flavor family, which offer different phenomenological consequences, from quark mass hierarchy to dark matter interactions with SM fields \cite{martinez2014some,martinez2014scalar,martinez2015scalar,martinez2015spin,MantillaDic2016}. 

The main goal of this article is to obtain predictable mass structures and parameters from the neutrino oscillation data by introducing a family non-universal and anomaly free $\mathrm{U(1)}'$ extension with three Higgs doublets and one Higgs singlet ($3$HD$+1$HS). The three doublets will generate the electroweak symmetry breaking, while the singlet induce the $U(1)'$ breaking spontaneously. The fermion and gauge sectors are also extended by new extra quarks and leptons (including Majorana fermions) with a $\mathbb{Z}_{2}$ symmetry and the $Z'$ gauge boson with non-universal interactions. A scalar singlet without vacuum expectation value is also included. The section \ref{sect:model} presents the model, its particle content and the Yukawa Lagrangian. 
The bosonic sector of the model is presented in the sections \ref{sect:Gauge} and \ref{sect:Higgs}. In section \ref{sect:Fermion-masses}, the mass expressions for all fermions are obtained as well as their mixing angles. Then, the section \ref{sect:PMNS-matrix} presents a procedure to test the consistency of the model with current neutrino oscillation data by fitting the Yukawa coupling constants of the neutrino Yukawa Lagrangian, and the section \ref{sect:HLFV} explores how much adecquate is the model in studying Lepton Flavor Violation (LFV) in Higgs decays. Finally, a discussion about the main results and some conclusions are outlined in the section \ref{sect:Conclusion}. 

\section{Some remarks of the model}
\label{sect:model}

\begin{table*}
\centering
\begin{tabular}{cc|cc|cc}
\hline\hline
Bosons	&	$X^{\pm}$	&	Quarks	&	$X^{\pm}$	&	Leptons	&	$X^{\pm}$	\\ \hline 
\multicolumn{2}{c}{Scalar Doublets}	&
\multicolumn{4}{c}{SM Fermionic Doublets}	\\ \hline\hline
$\Phi_{1}=\left(\begin{array}{c}
\phi_{1}^{+} \\ \frac{h_{1}+v_{1}+i\eta_{1}}{\sqrt{2}}	
\end{array}\right)$	&	$\sfrac{+2}{3}^{+}$	&
$q^{1}_{L}=\left(\begin{array}{c}u^{1}	\\ d^{1} \end{array}\right)_{L}$
	&	$\sfrac{+1}{3}^{+}$	&	
$\ell^{e}_{L}=\left(\begin{array}{c}\nu^{e} \\ e^{e} \end{array}\right)_{L}$
	&	$0^{+}$	\\
$\Phi_{2}=\left(\begin{array}{c}
\phi_{2}^{+} \\ \frac{h_{2}+v_{2}+i\eta_{2}}{\sqrt{2}}	
\end{array}\right)$	&	$\sfrac{+1}{3}^{-}$	&
$q^{2}_{L}=\left(\begin{array}{c}u^{2} \\ d^{2} \end{array}\right)_{L}$
	&	$0^{-}$	&
$\ell^{\mu}_{L}=\left(\begin{array}{c}\nu^{\mu} \\ e^{\mu} \end{array}\right)_{L}$
	&	$0^{+}$		\\
$\Phi_{3}=\left(\begin{array}{c}
\phi_{3}^{+} \\ \frac{h_{3}+v_{3}+i\eta_{3}}{\sqrt{2}}	
\end{array}\right)$	&	$\sfrac{+1}{3}^{+}$	&
$q^{3}_{L}=\left(\begin{array}{c}u^{3} \\ d^{3} \end{array}\right)_{L}$
	&	$0^{+}$	&
$\ell^{\tau}_{L}=\left(\begin{array}{c}\nu^{\tau} \\ e^{\tau} \end{array}\right)_{L}$
	&	$-1^{+}$	\\   \hline\hline
	
\multicolumn{2}{c}{Scalar Singlets} &\multicolumn{4}{c}{SM Fermionic Singlets}	\\ \hline\hline
\begin{tabular}{c}$\chi  =\frac{\xi_{\chi}  +v_{\chi}  +i\zeta_{\chi}}{\sqrt{2}}$\\$\sigma$\end{tabular}	&
\begin{tabular}{c}$\sfrac{-1}{3}^{+}$\\$\sfrac{-1}{3}^{-}$\end{tabular}	&
\begin{tabular}{c}$u_{R}^{1,3}$\\$u_{R}^{2}$\\$d_{R}^{1,2,3}$\end{tabular}	&	 
\begin{tabular}{c}$\sfrac{+2}{3}^{+}$\\$\sfrac{+2}{3}^{-}$\\$\sfrac{-1}{3}^{-}$\end{tabular}	&
\begin{tabular}{c}$e_{R}^{e}$\\$e_{R}^{\mu}$\\$e_{R}^{\tau}$\end{tabular}	&	
\begin{tabular}{c}$\sfrac{-4}{3}^{+}$\\$\sfrac{-1}{3}^{+}$\\$\sfrac{-4}{3}^{-}$\end{tabular}\\   \hline \hline 

\multicolumn{2}{c}{Gauge bosons}&	\multicolumn{2}{c}{Non-SM Quarks}&	\multicolumn{2}{c}{Non-SM Leptons}\\ \hline \hline
\begin{tabular}{c}$W^{\pm}_{\mu}$\\$W^{3}_{\mu}$\end{tabular}	&
\begin{tabular}{c}$0^{+}$\\$0^{+}$\end{tabular}	&
\begin{tabular}{c}$\mathcal{T}_{L}$\\$\mathcal{T}_{R}$\end{tabular}	&
\begin{tabular}{c}$\sfrac{+1}{3}^{-}$\\$\sfrac{+2}{3}^{-}$\end{tabular}	&
\begin{tabular}{c}$\nu_{R}^{e,\mu,\tau}$\\$\mathcal{N}_{R}^{e,\mu,\tau}$\end{tabular} 	&	
\begin{tabular}{c}$\sfrac{+1}{3}^{+}$\\$0^{+}$\end{tabular}\\
$B_{\mu}$	&	  $0^{+}$	&	
$\mathcal{J}^{1,2}_{L}$	&	  $0^{+}$	&
$\mathcal{E}_{L}^{1},\mathcal{E}_{R}^{2}$	&	$-1^{+}$	\\

$\Xi_{\mu}$	&	 $0^{+}$	&
$\mathcal{J}^{1,2}_{R}$	&	 $\sfrac{-1}{3}^{+}$	&
$\mathcal{E}_{R}^{1},\mathcal{E}_{L}^{2}$	&	$\sfrac{-2}{3}^{+}$	\\ \hline \hline
\end{tabular}
\caption{Non-universal $X$ quantum number and $\mathbb{Z}_{2}$ parity for SM and non-SM fermions.}
\label{tab:Particle-content}
\end{table*}

The model proposes the existence of a non-universal gauge group $\mathrm{\mathrm{U}(1)}_{X}$ whose gauge boson and coupling constant are $Z_{\mu}'$ and $g_{X}$, respectively. This additional gauge symmetry introduces new triangle chiral anomaly equations
\begin{eqnarray}
\label{eq:Chiral-anomalies}
\left[\mathrm{\mathrm{SU}(3)}_{C} \right]^{2} \mathrm{\mathrm{U}(1)}_{X} \rightarrow & A_{C} &= \sum_{Q}X_{Q_{L}} - \sum_{Q}X_{Q_{R}}	\nonumber	\\
\left[\mathrm{\mathrm{SU}(2)}_{L} \right]^{2} \mathrm{\mathrm{U}(1)}_{X} \rightarrow & A_{L}  &= \sum_{\ell}X_{\ell_{L}} + 3\sum_{Q}X_{Q_{L}}	\nonumber	\\
\left[\mathrm{\mathrm{U}(1)}_{Y} \right]^{2}   \mathrm{\mathrm{U}(1)}_{X} \rightarrow & A_{Y^{2}}&=
	\sum_{\ell, Q}\left[Y_{\ell_{L}}^{2}X_{\ell_{L}}+3Y_{Q_{L}}^{2}X_{Q_{L}} \right]\nonumber	\\ &&- \sum_{\ell,Q}\left[Y_{\ell_{R}}^{2}X_{L_{R}}+3Y_{Q_{R}}^{2}X_{Q_{R}} \right]	\nonumber	\\
\mathrm{\mathrm{U}(1)}_{Y}   \left[\mathrm{\mathrm{U}(1)}_{X} \right]^{2} \rightarrow & A_{Y}&=
	\sum_{\ell, Q}\left[Y_{\ell_{L}}X_{\ell_{L}}^{2}+3Y_{Q_{L}}X_{Q_{L}}^{2} \right]\nonumber	\\ &&- \sum_{\ell, Q}\left[Y_{\ell_{R}}X_{\ell_{R}}^{2}+3Y_{Q_{R}}X_{Q_{R}}^{2} \right]	\nonumber	\\
\left[\mathrm{\mathrm{U}(1)}_{X} \right]^{3} \rightarrow & A_{X}&=
	\sum_{\ell, Q}\left[X_{\ell_{L}}^{3}+3X_{Q_{L}}^{3} \right]\nonumber	\\ &&- \sum_{\ell, Q}\left[X_{\ell_{R}}^{3}+3X_{Q_{R}}^{3} \right]		\nonumber	\\
\left[\mathrm{Grav} \right]^{2}   \mathrm{\mathrm{U}(1)}_{X} \rightarrow & A_{\mathrm{G}}&=
	\sum_{\ell, Q}\left[X_{\ell_{L}}+3X_{Q_{L}} \right]\\ &&- \sum_{\ell, Q}\left[X_{\ell_{R}}+3X_{Q_{R}} \right]\nonumber		
\end{eqnarray}
which can be solved by assigning non-trivial $X$-quantum numbers to the fermions of the SM \cite{martinez2014some,MantillaDic2016}. Non-universal solutions emerge naturally if new quarks and leptons are added with the condition that they must acquire masses at a larger scale than the electroweak ones. All the new particles are assumed to be singlets under the gauge $SU(2)_L$ group. So, they acquire masses by the VEV of a new Higgs singlet (1HS) field, $\chi$, which has $\mathrm{\mathrm{U}(1)}_{X}$ charge ($X=-1/3$) 
 in such a way that it spontaneously breaks the new gauge symmetry. Another scalar singlet $\sigma$ identical to $\chi$ but without VEV is introduced \cite{martinez2014scalar,martinez2015scalar,martinez2015spin}. 

On the other hand, the fermion mass hierarchy can be understood with a three Higgs doublet model (3HD), with vacuum expectation values (VEV) $v_{1}>v_{2}>v_{3}$ and the general constraint $v^2=v_1^2+v_2^2+v_3^2$, with $v=246$ GeV the electroweak breaking scale. The first VEV, $v_{1}$, can be associated with the mass of the top quark $t$ at $10^2$ GeV. Second, the tau lepton $\tau$ and the bottom quark $b$ might acquire mass through $v_{2}$ at $1$ GeV. Third, the muon $\mu$ and the strange quark $s$ may get mass by $v_{3}$ at $10^{2}$ MeV scale. Regarding to the charm quark $c$, and the complete first generation of charged fermions $(u,d,e)$ could get masses through see-saw mechanisms and radiative corrections so as their masses get smaller without needing unpleasant fine-tunnings. Thus, the combination among the 3HD, the 1HS, and the requirement of specific  $\mathbb{Z}_{2}$ transformations lead us to predictable mass structures of the fermions. 
The chosen particle spectrum is presented in the table \ref{tab:Particle-content} where three new quarks ($\mathcal{T}$, $\mathcal{J}^{1,2}$), two charged leptons ($\mathcal{E}^{1,2}$) and three right-handed neutrinos $\nu_{R}^{e,\mu,\tau}$ were introduced such that the model is free from chiral anomalies. By replacing the $X$-charges shown in table \ref{tab:Particle-content} in the equations \eqref{eq:Chiral-anomalies}, the complete set of chiral anomalies get cancelled indentically
\begin{equation}
\begin{split}
A_{C} =& 1 - 1,	\\
A_{L}  =& -2 + 3(\sfrac{2}{3}),	\\
A_{Y^{2}}=&
	\left[\sfrac{-26}{ 3}+3(\sfrac{2}{3}) \right] - \left[\sfrac{- 56}{ 3}+3(4) \right],	\\
A_{Y}=&
	\left[\sfrac{-44}{ 9}+3(\sfrac{2}{9}) \right] - \left[\sfrac{- 92}{ 9}+3(2) \right],	\\
A_{X}=&
	\left[\sfrac{-89}{27}+3(\sfrac{1}{9}) \right] - \left[\sfrac{-161}{27}+3(1) \right],	\\
A_{\mathrm{G}}=&
	\left[\sfrac{-11}{ 3}+3(3) \right] - \left[\sfrac{-11}{3}+3(3) \right].
\end{split}
\end{equation}
The $\mathbb{Z}_{2}$ parities are also shown as superscripts in the $X$-charges. It is to note that, despite the scalar doublets $\Phi_{2}$ and $\Phi_{3}$ have the same $X$ charge, they have opposite $\mathbb{Z}_{2}$ parity such that their couplings to fermions are complementary. 

The addition of $\nu_{R}$ 
allow the coupling with $\ell_{L}$ through $\Phi_{3}$, generating mass terms to the active neutrinos. However, the experiments suggest us that their masses are smaller than their charged lepton partners in many order of magnitudes. This huge difference could be explained by the well-known inverse SSM which is implemented here by introducing three Majorana fermions, $\mathcal{N}_{R}$, which couple to $\nu_{R}$ via the scalar singlet $\chi$. The existence of the corresponding Majorana mass term induces the inverse SSM yielding to three light and three 
quasi-degenerated heavy neutrinos at the TeV scale. One important consequence of the smallness of $v_{3}$ is that the mass of the Majorana neutrinos can be as low as the MeV scale.  

The Yukawa Lagrangian of the model for the up-like, down-like, neutral and charged fermions are respetively:
\begin{equation}
\begin{split}
-\mathcal{L}_{U} &= 
h_{3 u}^{1 1}\overline{q_{L}^{1}}\tilde{\Phi}_{3}u_{R}^{1} + 
h_{2 u}^{1 2}\overline{q_{L}^{1}}\tilde{\Phi}_{2}u_{R}^{2} + 
h_{3 u}^{1 3}\overline{q_{L}^{1}}\tilde{\Phi}_{3}u_{R}^{3} \\ &+ 
h_{1 u}^{2 2}\overline{q_{L}^{2}}\tilde{\Phi}_{1}u_{R}^{2} + 
h_{1 u}^{3 1}\overline{q_{L}^{3}}\tilde{\Phi}_{1}u_{R}^{1} + 
h_{1 u}^{3 3}\overline{q_{L}^{3}}\tilde{\Phi}_{1}u_{R}^{3} \\ &+
h_{2 \mathcal{T}}^{1} \overline{q_{L}^{1}}\tilde{\Phi}_{2}\mathcal{T}_{R} +
h_{1 \mathcal{T}}^{2} \overline{q_{L}^{2}}\tilde{\Phi}_{1}\mathcal{T}_{R} +
g_{\sigma u}^{1}\overline{\mathcal{T}_{L}}\sigma u_{R}^{1} \\ &+ 
g_{\chi u}^{2}\overline{\mathcal{T}_{L}}\chi u_{R}^{2} \,+ 
g_{\sigma u}^{3}\overline{\mathcal{T}_{L}}\sigma u_{R}^{3} \,+ 
g_{\chi \mathcal{T}}\overline{\mathcal{T}_{L}}\chi \mathcal{T}_{R} + \mathrm{h.c.},
\end{split}
\label{eq:Up-Lagrangian}
\end{equation}
\begin{equation}
\begin{split}
-\mathcal{L}_{D} &= 
h_{1 \mathcal{J}}^{1 1}\overline{q_{L}^{1}}{\Phi}_{1}\mathcal{J}_{R}^{1} + 
h_{2 \mathcal{J}}^{2 1}\overline{q_{L}^{2}}{\Phi}_{2}\mathcal{J}_{R}^{1} + 
h_{3 \mathcal{J}}^{3 1}\overline{q_{L}^{3}}{\Phi}_{3}\mathcal{J}_{R}^{1} \\ &+
h_{1 \mathcal{J}}^{1 2}\overline{q_{L}^{1}}{\Phi}_{1}\mathcal{J}_{R}^{2} + 
h_{2 \mathcal{J}}^{2 2}\overline{q_{L}^{2}}{\Phi}_{2}\mathcal{J}_{R}^{2} + 
h_{3 \mathcal{J}}^{3 2}\overline{q_{L}^{3}}{\Phi}_{3}\mathcal{J}_{R}^{2} \\ &+ 
h_{3 d}^{2 1}\overline{q_{L}^{2}}{\Phi}_{3}d_{R}^{1} +
h_{3 d}^{2 2}\overline{q_{L}^{2}}{\Phi}_{3}d_{R}^{2} +
h_{3 d}^{2 3}\overline{q_{L}^{2}}{\Phi}_{3}d_{R}^{3} \\ &+
h_{2 d}^{3 1}\overline{q_{L}^{3}}{\Phi}_{2}d_{R}^{1} +
h_{2 d}^{3 2}\overline{q_{L}^{3}}{\Phi}_{2}d_{R}^{2} +
h_{2 d}^{3 3}\overline{q_{L}^{3}}{\Phi}_{2}d_{R}^{3} \\ &+ 
g_{\sigma d}^{1 1}\overline{\mathcal{J}_{L}^{1}}\sigma^{*} d_{R}^{1} + 
g_{\sigma d}^{1 1}\overline{\mathcal{J}_{L}^{1}}\sigma^{*} d_{R}^{2} +
g_{\sigma d}^{1 3}\overline{\mathcal{J}_{L}^{1}}\sigma^{*} d_{R}^{3} \\ &+ 
g_{\sigma d}^{2 1}\overline{\mathcal{J}_{L}^{2}}\sigma^{*} d_{R}^{1} +
g_{\sigma d}^{2 2}\overline{\mathcal{J}_{L}^{2}}\sigma^{*} d_{R}^{2} +
g_{\sigma d}^{2 3}\overline{\mathcal{J}_{L}^{2}}\sigma^{*} d_{R}^{3} \\ &+ 
g_{\chi \mathcal{J}}^{1}\overline{\mathcal{J}_{L}^{1}}\chi^{*} \mathcal{J}_{R}^{1} + 
g_{\chi \mathcal{J}}^{2}\overline{\mathcal{J}_{L}^{2}}\chi^{*} \mathcal{J}_{R}^{2} + \mathrm{h.c.},
\end{split}
\label{eq:Down-Lagrangian}
\end{equation}
\begin{equation}
\begin{split}
-\mathcal{L}_{N} &=
h_{3 \nu}^{e e}\overline{\ell^{e}_{L}}\tilde{\Phi}_{3}\nu^{e}_{R} + 
h_{3 \nu}^{e \mu}\overline{\ell^{e}_{L}}\tilde{\Phi}_{3}\nu^{\mu}_{R} + 
h_{3 \nu}^{e \tau}\overline{\ell^{e}_{L}}\tilde{\Phi}_{3}\nu^{\tau}_{R} \\ &+
h_{3 \nu}^{\mu e}\overline{\ell^{\mu}_{L}}\tilde{\Phi}_{3}\nu^{e}_{R} +
h_{3 \nu}^{\mu \mu}\overline{\ell^{\mu}_{L}}\tilde{\Phi}_{3}\nu^{\mu}_{R} + 
h_{3 \nu}^{\mu \tau}\overline{\ell^{\mu}_{L}}\tilde{\Phi}_{3}\nu^{\tau}_{R} \\ &+
g_{\chi \mathcal{N}}^{i j} \overline{\nu_{R}^{i\;C}} \chi^{*} \mathcal{N}_{R}^{j} +
\frac{1}{2} \overline{\mathcal{N}_{R}^{i\;C}} M^{ij}_{\mathcal{N}} \mathcal{N}_{R}^{j} + \mathrm{h.c.},
\end{split}
\label{eq:Neutrino-Lagrangian}
\end{equation}
\begin{equation}
\begin{split}
-\mathcal{L}_{E} &=
h_{3 e}^{e \mu}\overline{\ell^{e}_{L}}\Phi_{3}e^{\mu}_{R} + 
h_{3 e}^{\mu \mu}\overline{\ell^{\mu}_{L}}\Phi_{3}e^{\mu}_{R} + 
h_{3 e}^{\tau e}\overline{\ell^{\tau}_{L}}\Phi_{3}e^{e}_{R} \\ &+
h_{2 e}^{\tau \tau}\overline{\ell^{\tau}_{L}}\Phi_{2}e^{\tau}_{R} +
h_{1 E}^{e 1}\overline{\ell^{e}_{L}}\Phi_{1}\mathcal{E}_{R}^{1} + 
h_{1 \mathcal{E}}^{\mu 1}\overline{\ell^{\mu}_{L}}\Phi_{1}\mathcal{E}_{R}^{1} \\ &+
g_{\chi e}^{1 e}\overline{\mathcal{E}_{L}^{1}}\chi^{*} e^{e}_{R} + 
g_{\chi e}^{2 \mu}\overline{\mathcal{E}_{L}^{2}}\chi e^{\mu}_{R} +
g_{\chi \mathcal{E}}^{1}\overline{\mathcal{E}_{L}^{1}}\chi \mathcal{E}_{R}^{1}  \\ &+ 
g_{\chi \mathcal{E}}^{2}\overline{\mathcal{E}_{L}^{2}}\chi^{*} \mathcal{E}_{R}^{2} + \mathrm{h.c.},
\end{split}
\label{eq:Electron-Lagrangian}
\end{equation}
where $\widetilde{\Phi}=i\sigma_2 \Phi^*$ are the scalar doublet conjugates and the Majorana mass components are denoted as $M^{ij}_{\mathcal{N}}$. The next section presents the acquisition of masses in the fermion sectors.

\section{Gauge bosons and masses}
\label{sect:Gauge}
The gauge bosons of the model comprises the vector sector of the SM plus the additional $\Xi_{\mu}$ gauge boson of the abelian extension $\mathrm{U(1)}_{X}$. The Gauge Lagrangian is
\begin{align}
\mathcal{L}_{\mathrm{Gauge}} = -\frac{1}{4}\mathrm{Tr}\left(\mathbf{W}^{\mu\nu}\mathbf{W}_{\mu\nu}\right) - \frac{1}{4}B^{\mu\nu}B_{\mu\nu} - \frac{1}{4}{\Xi}^{\mu\nu}\Xi_{\mu\nu}
\end{align}
where $\Xi_{\mu\nu}$ is the strength-field tensor of the $\Xi_{\mu}$ gauge boson
\begin{equation}
\begin{split}
\Xi_{\mu\nu} = \partial_{\mu}\Xi_{\nu} - \partial_{\nu}\Xi_{\mu}.
\end{split}
\end{equation}

The gauge boson masses, on the other hand, come from the kinetic part of the Higgs Lagrangian
\begin{equation}
\begin{split}
\mathcal{L}_{\mathrm{Higgs}}^{\mathrm{Kin}} &=
 \frac{1}{2}\sum_{1,2,3}\left(D^{\mu}\Phi_{i}\right)^{\dagger}\left(D_{\mu}\Phi_{i}\right)
+\frac{1}{2}\left(D^{\mu}\chi\right)^{*}\left(D_{\mu}\chi\right)
,
\end{split}
\end{equation}
where de covariant derivatives are
\begin{subequations}
\begin{align}
D_{\mu}\Phi_{i} &= \partial_{\mu}\Phi_{i} - ig \mathbf{W}_{\mu}\Phi_{i}
 - ig'Y B_{\mu}\Phi_{i} - ig_{X}X_{i} \Xi_{\mu}\Phi_{i},\\
D_{\mu}\chi &= \partial_{\mu}\chi + \tfrac{i g_{X}}{3} \Xi_{\mu}\chi.
\end{align}
\end{subequations}
By evaluating the Higgs fields at their VEVs the gauge boson masses appear. The mass of the $W^{\pm}_{\mu}$ is
\begin{equation}
m_{W}^{2}=\frac{g^{2}}{4}\left(v_{1}^{2}+v_{2}^{2}+v_{3}^{2}\right)=\frac{g^{2}v^{2}}{4}
\end{equation}
where $v$ is the complete electroweak VEV. Regarding to the neutral gauge bosons, the mass matrix in the basis $\mathbf{W}^{0}_{\mu}=(B_{\mu},W_{\mu}^{3},\Xi_{\mu})$ is
\begin{align}
M_{{W}^{0}}^{2} \approx 
\frac{1}{4}
\begin{pmatrix}
	             g'^2 v^2	&
	           	-g g' v^2	&
	 \frac{4}{3} g' g_X v^2 \\
	            -g  g'  v^2	&	
	             g^2 v^2	&
	-\frac{4}{3} g  g_X v^2 \\
	 \frac{4}{3} g' g_X v^2	&
	-\frac{4}{3} g  g_X v^2	&
	 \frac{4}{9} g_X^2 v_{\chi}^2
\end{pmatrix}.
\end{align}
Its determinant is null as it is hoped because the existence of a massless gauge boson, the photon $A_{\mu}$. In addition, there are two massive gauge bosons, the electroweak $Z_{\mu}$ at GeV scale, and the new $Z'_{\mu}$ at TeV
\begin{align}
m_{Z }^{2} \approx \frac{g^{2}+{g'}^{2}}{4}v^{2} = \frac{g^{2}v^{2}}{4c_{W}^{2}},\qquad 
m_{Z'}^{2} \approx \frac{g_{X}^2 v_{\chi}^2}{9}
\end{align}
The mass eigenstates $\mathbf{Z}_{\mu}=(A_{\mu},Z_{\mu},Z'_{\mu})$ are obained as $\mathbf{Z}_{\mu}=R_{{W}^{0}}\mathbf{W}^{0}_{\mu}$ through the mixing matrix $R_{{W}^{0}}$. In the CKM-parametrization its angles are
\begin{subequations}
\begin{align}
\tan \theta_{12}^{{W}^{0}} = \frac{g'}{g},\;\; \tan \theta_{13}^{{W}^{0}}=0,\;\;
\tan \theta_{23}^{{W}^{0}} \approx \frac{3 g v^2}{g_X c_W v_{\chi }^2},
\end{align}
\end{subequations}
and the first angle turns out to be the well-known Weinberg angle $\tan \theta_{12}^{{W}^{0}}=t_{W}$.

\section{Higgs potential and scalar masses}
\label{sect:Higgs}
The scalar potential of the model is stablished according to the $\mathrm{U(1)}_{X}$ charges and $\mathbb{Z}_{2}$ parities shown in the table \ref{tab:Particle-content}. So, the most general potential invariant under the $\mathrm{G_{SM}}\otimes \mathrm{U(1)}_{X}\otimes \mathbb{Z}_{2}$ symmetry is
\begin{widetext}
\begin{equation}
\begin{split}	
V_{\mathrm{H}} &= \mu_{1}^{2}\Phi_{1}^{\dagger}\Phi_{1} + \mu_{2}^{2}\Phi_{2}^{\dagger}\Phi_{2}
 + \mu_{3}^{2}\Phi_{3}^{\dagger}\Phi_{3} + \mu_{\chi}^{2}\chi^{*}\chi
 + \lambda_{\chi\chi}\left(\chi^{*}\chi\right)^{2}	\\ &
 - \tfrac{f_{2}}{\sqrt{2}}\left(\Phi_{1}^{\dagger}\Phi_{2}\chi + \mathrm{h.c.} \right)
 - \tfrac{f_{3}}{\sqrt{2}}\left(\Phi_{1}^{\dagger}\Phi_{3}\chi + \mathrm{h.c.} \right) \\ &
 + \lambda_{11}\left(\Phi_{1}^{\dagger}\Phi_{1}\right)^{2} 
 + \lambda_{12}\left(\Phi_{1}^{\dagger}\Phi_{1}\right)\left(\Phi_{2}^{\dagger}\Phi_{2}\right) 
 -\lambda_{12}'\left(\Phi_{1}^{\dagger}\Phi_{2}\right)\left(\Phi_{2}^{\dagger}\Phi_{1}\right)  \\ &
 + \lambda_{22}\left(\Phi_{2}^{\dagger}\Phi_{2}\right)^{2} 
 + \lambda_{23}\left(\Phi_{2}^{\dagger}\Phi_{2}\right)\left(\Phi_{3}^{\dagger}\Phi_{3}\right) 
 -\lambda_{23}'\left(\Phi_{2}^{\dagger}\Phi_{3}\right)\left(\Phi_{3}^{\dagger}\Phi_{2}\right) \\ &
 + \lambda_{33}\left(\Phi_{3}^{\dagger}\Phi_{3}\right)^{2} 
 + \lambda_{13}\left(\Phi_{1}^{\dagger}\Phi_{3}\right)\left(\Phi_{3}^{\dagger}\Phi_{1}\right)
 -\lambda_{13}'\left(\Phi_{1}^{\dagger}\Phi_{3}\right)\left(\Phi_{3}^{\dagger}\Phi_{1}\right) \\ &
 + \lambda_{1\chi}\left(\Phi_{1}^{\dagger}\Phi_{1}\right)\left(\chi^{*}\chi \right)
 + \lambda_{2\chi}\left(\Phi_{2}^{\dagger}\Phi_{2}\right)\left(\chi^{*}\chi \right)
 + \lambda_{3\chi}\left(\Phi_{3}^{\dagger}\Phi_{3}\right)\left(\chi^{*}\chi \right).
\end{split}
\end{equation}
\end{widetext}

\subsection{Minimization of the potential}
The previous potential is minimized by differentiating it respect to each one of the VEVs and isolating the quadratic constants $\mu_{\alpha}$ where $\alpha,\beta=1,2,3,\chi$. Thus, the following constants are obtained
\begin{subequations}
\begin{align}
-\mu_{1}^{2}&=\sum_{\alpha=1}^{\chi}{ {\Lambda_{1\alpha}} {v_{\alpha}^{2}}}
 - \frac{ {v_{\chi}}{v_{2}}{f_{2}}+{v_{\chi}}{v_{3}}{f_{3}}}{2  {v_{1}}},
\\
-\mu_{2}^{2}&=\sum_{\alpha=1}^{\chi}{ {\Lambda_{2\alpha}} {v_{\alpha}^{2}}}
 - \frac{ {f_{2}}{v_{1}}{v_{\chi}}}{2{v_{2}}},
\\
-\mu_{3}^{2}&=\sum_{\alpha=1}^{\chi}{ {\Lambda_{3\alpha}} {v_{\alpha}^{2}}}
 - \frac{ {f_{3}}{v_{1}}{v_{\chi}}}{2  {v_{3}}},
\\
-\mu_{\chi}^{2}&=\sum_{\alpha=1}^{\chi}{ {\Lambda_{\chi \alpha}} {v_{\alpha}^{2}}}
 - \frac{ {v_{1}}{v_{2}}{f_{2}}+{v_{1}}{v_{3}}{f_{3}} }{2  {v_{\chi}}}
\end{align}
\end{subequations}
where the constants $\Lambda_{\alpha\beta}=\Lambda_{\beta\alpha}$ are ($i,j=1,2,3$)
\begin{equation}
\left\lbrace
\begin{split}
\Lambda_{\alpha\alpha}&=\lambda_{\alpha\alpha},\\
\Lambda_{ij}&=(\lambda_{ij}-\lambda_{ij}')/2\\
\Lambda_{i\chi}&= \lambda_{i\chi}/2.
\end{split}
\right .
\end{equation}

\subsection{Charged scalar boson masses}
The mass matrix of the charged bosons is obtained by calculating the Hessian matrix respect to the charged components of the Higgs doublets. In the basis $\boldsymbol{\phi^{\pm}}=(\phi_{1}^{\pm},\phi_{2}^{\pm},\phi_{3}^{\pm})$ it turns out to be
\begin{equation}
M_{\mathrm{C}}^{2}\approx
\frac{1}{4}
\begin{pmatrix}
	 \dfrac{v_i f_{i} v_{\chi }}{v_1} & 
	-f_{2} v_{\chi }	& 
	-f_{3} v_{\chi }	\\
	-f_{2} v_{\chi }	& 
	\dfrac{f_{2} v_{\chi } v_1}{v_2} & 
	0	\\
	-f_{3} v_{\chi }	& 
	0	& 
	\dfrac{f_{3 } v_{\chi } v_1}{v_3}
\end{pmatrix}
\end{equation}
where $v_i f_{i}=v_2 f_{2}+v_3 f_{3}$. Its determinant is null as it is hoped because the existence of $G_{W}^{\pm}$, the Goldstone bosons of $W^{\pm}_{\mu}$. Additionally there exist two physical charged bosons $H^{\pm}_{1}$ and $H^{\pm}_{2}$ which acquire mass at TeV scale.

The masses of the physical charged bosons are
\begin{eqnarray}
&m_{H^{\pm}_{1,2}}^{2}&\approx 
\frac{{f_{2} } \left(v_1^2+v_2^2\right) {v_{\chi} }}{8 v_1 v_2}+\frac{{f_{3} } \left(v_1^2+v_3^2\right) {v_{\chi} }}{8 v_1 v_3}\\&
\pm & \sqrt{\tfrac{f_{2} ^2 \left(v_1^2+v_2^2\right){}^2 v_{\chi}^{2}}{64 v_1^2 v_2^2}-\tfrac{{f_{2} } {f_{3} } \left(v_1^4-v_2^2 v_3^2\right) v_{\chi}^{2}}{32 v_1^{2} v_2 v_3}+\tfrac{f_{3} ^2 \left(v_1^2+v_3^2\right){}^2 v_{\chi}^{2}}{64 v_1^2 v_3^2}}\nonumber .
\end{eqnarray}
The mixing matrix $R_{\mathrm{C}}$ diagonalizes the mass matrix $M_{\mathrm{C}}^{2}$ obtaining the mass eigenstates $\mathbf{H}^{\pm}=R_{\mathrm{C}}\boldsymbol{\phi^{\pm}}$ which are expressed in the basis $\mathbf{H}^{\pm}=(G_{W}^{\pm},H_{1}^{\pm},H_{2}^{\pm})$. Its corresponding mixing angles in the CKM parametrization are
\begin{align}
\tan^{2} \theta_{12}^{C} =\frac{v_{2}^{2}}{v_{1}^{2}},
\quad \tan^{2} \theta_{13}^{C} =\frac{v_{3}^{2}}{v_{1}^{2}+v_{2}^{2}},
\quad \tan^{2} \theta_{23}^{\mathrm{C}} \approx 0	
\end{align}

\subsection{CP-odd boson masses}
The mass matrix of the CP-odd (pseudoscalar) bosons is obtained by calculating the Hessian matrix respect to the CP-odd components of the Higgs doublets. In the basis $\boldsymbol{\eta}=(\eta_{1},\eta_{2},\eta_{3},\zeta_{\chi})$ it turns out to be
\begin{equation}
M_{\mathrm{odd}}^{2} = \frac{1}{4}
\begin{pmatrix}
	 \dfrac{v_i f_{i} v_{\chi }}{v_1} & 
	-f_{2} v_{\chi }	& 
	-f_{3} v_{\chi }	& 
	-f_{i} v_i	\\
	-f_{2} v_{\chi }	& 
	\dfrac{f_{2} v_{\chi } v_1}{v_2} & 
	0 & 
	f_{2}v_{1}	\\
	-f_{3 } v_{\chi }	& 
	0	& 
	\dfrac{f_{3 } v_{\chi } v_1}{v_3}	&
	 f_{3} v_{1}	\\
	-f_{i} v_i 	&
	 f_{2} v_{1}	&
	 f_{3} v_{1}	&
	\dfrac{v_i f_{i} v_{1}}{v_{\chi}}
\end{pmatrix}
\end{equation}
where $v_i f_{i}=v_2 f_{2}+v_3 f_{3}$. Its determinant is null as it is hoped because the existence of $G_{Z}$ and $G_{Z'}$, the Goldstone bosons of $Z_{\mu}$ and $Z'_{\mu}$, respectively. Additionally there exist two physical pseudoscalar bosons $A_{1}$ and $A_{2}$ which acquire mass at TeV.

The masses of the physical pseudoscalar bosons are
\begin{eqnarray}
&m_{A_{1,2}}^{2}&\approx
\frac{{f_{2} } \left(v_1^2+v_2^2\right) {v_{\chi} }}{8 v_1 v_2}+\frac{{f_{3} } \left(v_1^2+v_3^2\right) {v_{\chi} }}{8 v_1 v_3}\\&
\pm & \sqrt{\tfrac{f_{2} ^2 \left(v_1^2+v_2^2\right){}^2 v_{\chi}^{2}}{64 v_1^2 v_2^2}-\tfrac{{f_{2} } {f_{3} } \left(v_1^4-v_2^2 v_3^2\right) v_{\chi}^{2}}{32 v_1^{2} v_2 v_3}+\tfrac{f_{3} ^2 \left(v_1^2+v_3^2\right){}^2 v_{\chi}^{2}}{64 v_1^2 v_3^2}}\nonumber ,
\end{eqnarray}
which are equal to the the charged bosons masses at $\mathcal{O}(v^{2})$. 
The mixing matrix $R_{\mathrm{odd}}$ diagonalizes the mass matrix $M_{\mathrm{odd}}^{2}$ obtaining the mass eigenstates $\mathbf{A}=R_{\mathrm{odd}}\boldsymbol{\eta}$ which are expressed in the basis $\mathbf{A}=(G_{Z},A_{1},A_{2},G_{Z'})$. Moreover, the diagonalization in this case is a little more complicated because there are four bosons instead of three in comparison with the charged scalar boson sector. So, it was implemented an extended-CKM parametrization which includes mixings with a fourth component. 
Thereby, the corresponding mixing angles are
\begin{subequations}
\begin{align}
   \tan^{2} \theta_{12}^{A}&=\frac{v_{2}^{2}}{v_{1}^{2}c_{14}^{2}},
\qquad \tan^{2} \theta_{13}^{A}=\frac{v_{3}^{2}}{v_{1}^{2}c_{14}^{2}+v_{2}^{2}},
\\ \tan^{2} \theta_{23}^{A}&\approx 0,
\qquad \qquad \tan^{2} \theta_{14}^{A}=\frac{v_{1}}{v_{\chi}},
\end{align}
\end{subequations}
where $c_{14}^{2}=\cos^{2} \theta_{14}^{\mathrm{odd}}$.

\subsection{CP-even boson masses}
The mass matrix of the CP-even (true scalar) bosons is obtained by calculating the Hessian matrix respect to the CP-even components of the Higgs doublets. In the basis $\mathbf{h}=(h_{1},h_{2},h_{3},\xi_{\chi})$ the CP-even mass matrix is
\begin{equation}
M_{\mathrm{even}}^{2}=
\begin{pmatrix}
	\mathcal{M}_{hh}	&	\mathcal{M}_{h\xi}	\\
	\mathcal{M}_{h\xi}^{\mathrm{T}}	&	\mathcal{M}_{\xi\xi}
\end{pmatrix},
\end{equation}
where the blocks are defined as
\begin{subequations}
\begin{align}
\mathcal{M}_{hh}&=
\begin{pmatrix}
	\Lambda_{11} v_{1}^2	& 
	\Lambda_{12} v_{1}v_{2}	& 
	\Lambda_{13} v_{1}v_{3}	\\
	\Lambda_{12} v_{1}v_{2}	& 
	\Lambda_{22} v_{2}^{2}	& 
	\Lambda_{23} v_{2}v_{3}	\\
	\Lambda_{13} v_{1}v_{3}	& 
	\Lambda_{23} v_{2}v_{3}	& 
	\Lambda_{33} v_{3}^{2}
\end{pmatrix}+M_{\mathrm{C}}^{2}	\\
\mathcal{M}_{h\xi}&=
\begin{pmatrix}
	\Lambda_{1\chi}v_{1}v_{\chi}-\frac{f_{i}v_{i}}{4}	\\
	\Lambda_{2\chi}v_{2}v_{\chi}-\frac{f_{2}v_{1}}{4}	\\
	\Lambda_{3\chi}v_{3}v_{\chi}-\frac{f_{3}v_{1}}{4}
\end{pmatrix}	\\
\mathcal{M}_{\xi\xi}&=
	\Lambda_{\chi\chi}v_{\chi}^{2}+\dfrac{v_i f_{i} v_{1}}{4 v_{\chi}}
.	
\end{align}
\end{subequations}

The mixing matrix $R_{\mathrm{even}}$ which diagonalizes the mass matrix $M_{\mathrm{even}}^{2}$ gives the mass eigenstates $\mathbf{H}=R_{\mathrm{even}}\mathbf{h}$ which are expressed in the basis $\mathbf{H}=(h,H_{1},H_{2},\mathcal{H})$. Moreover, $R_{\mathrm{even}}$ splits in a see-saw rotation $R_{\mathrm{even}}^{\mathrm{SS}}$ and a block-diagonal rotation $R_{\mathrm{even}}^{B}$ such that $R_{\mathrm{even}}=R_{\mathrm{even}}^{\mathrm{B}}R_{\mathrm{even}}^{\mathrm{SS}}$. 

Since $|\mathcal{M}_{hh}|<|\mathcal{M}_{h\xi}|<|\mathcal{M}_{\xi\xi}|$ the see-saw procedure will be implemented by following the reference \cite{MantillaDic2016} which block-diagonalizes $\mathcal{M}_{hh}$ such that the $h$ scalars get separated from the $\xi$ ones. The following approximations are made on the blocks in order to avoid cumbersome expressions after rotating out the $\xi$ scalars: 
\begin{align}
\mathcal{M}_{h\xi} \approx 
\begin{pmatrix} 
	\Lambda_{1\chi}v_{1}v_{\chi}	\\
	\Lambda_{2\chi}v_{2}v_{\chi}	\\
	\Lambda_{3\chi}v_{3}v_{\chi}
\end{pmatrix},	\qquad 
\mathcal{M}_{\xi\xi} \approx 
\Lambda_{\chi\chi}v_{\chi}^{2}
.
\end{align}
The see-saw rotation $R_{\mathrm{even,SS}}$ and its angle $\Theta_{\mathrm{even}}$ are 
\begin{equation}
R_{\mathrm{even}}^{{\mathrm{SS}}} = 
\begin{pmatrix}
	1	&	-\Theta_{\mathrm{even}}^{\dagger}	\\
	\Theta_{\mathrm{even}}	&	1
\end{pmatrix},
\end{equation}
\begin{equation}
\Theta_{\mathrm{even}}^{\dagger} = \mathcal{M}_{\xi\xi}^{-1}\mathcal{M}_{h\xi}=
\begin{pmatrix} 
	\frac{\Lambda_{1\chi}v_{1}}{\Lambda_{\chi\chi}v_{\chi}}	\\
	\frac{\Lambda_{2\chi}v_{2}}{\Lambda_{\chi\chi}v_{\chi}}	\\
	\frac{\Lambda_{3\chi}v_{3}}{\Lambda_{\chi\chi}v_{\chi}}
\end{pmatrix}.
\end{equation}
The block-diagonalization acts in the following way
\begin{align}
R_{\mathrm{even}}^{{\mathrm{SS}}} \mathcal{M}_{hh} \left( R_{\mathrm{even}}^{{\mathrm{SS}}}\right)^{\mathrm{T}} = 
\begin{pmatrix}
	{M}_{hh}^{2}	&	0	\\
	0	&	{M}_{\xi\xi}^{2}
\end{pmatrix}.
\end{align}
where the new blocks are
\begin{align}
{M}_{hh}^{2} \approx \mathcal{M}_{hh}
 - \mathcal{M}_{h\xi}\mathcal{M}_{\xi\xi}^{-1}\mathcal{M}_{h\xi}^{\mathrm{T}},	\qquad
{M}_{\xi\xi}^{2} \approx \mathcal{M}_{\xi\xi}.
\end{align}

The resulting matrix $M_{hh}$ has the same algebraic structure of $\mathcal{M}_{hh}$ with new definitions of the constants $\Lambda_{ij}$'s, where $i,j=1,2,3$. The matrix turns out to be
\begin{align}
{M}_{hh}^{2} &\approx \mathcal{M}_{hh}
 - \mathcal{M}_{h\xi}\mathcal{M}_{\xi\xi}^{-1}\mathcal{M}_{h\xi}^{\mathrm{T}}	\\&\approx 
\begin{pmatrix}
	\widetilde{\Lambda}_{11} v_{1}^2	& 
	\widetilde{\Lambda}_{12} v_{1}v_{2}	& 
	\widetilde{\Lambda}_{13} v_{1}v_{3}	\\
	\widetilde{\Lambda}_{12} v_{1}v_{2}	& 
	\widetilde{\Lambda}_{22} v_{2}^{2}	& 
	\widetilde{\Lambda}_{23} v_{2}v_{3}	\\
	\widetilde{\Lambda}_{13} v_{1}v_{3}	& 
	\widetilde{\Lambda}_{23} v_{2}v_{3}	& 
	\widetilde{\Lambda}_{33} v_{3}^{2}
\end{pmatrix}+M_{\mathrm{C}}^{2}	\nonumber
\end{align}
where the tilde constants are
\begin{align}
\begin{split}
\widetilde{\Lambda}_{11} &= \Lambda_{11}-\frac{\Lambda_{1\chi}^2}{\Lambda_{\chi\chi}},	\qquad	
\widetilde{\Lambda}_{12}  = \Lambda_{12}-\frac{\Lambda_{1\chi} \Lambda_{2\chi}}{\Lambda_{\chi\chi}},	\\
\widetilde{\Lambda}_{22} &= \Lambda_{22}-\frac{\Lambda_{2\chi}^2}{\Lambda_{\chi\chi}},	\qquad
\widetilde{\Lambda}_{23}  = \Lambda_{23}-\frac{\Lambda_{2\chi} \Lambda_{3\chi}}{\Lambda_{\chi\chi}},	\\
\widetilde{\Lambda}_{33} &= \Lambda_{33}-\frac{\Lambda_{3\chi}^2}{\Lambda_{\chi\chi}},	\qquad 
\widetilde{\Lambda}_{13}  = \Lambda_{13}-\frac{\Lambda_{1\chi} \Lambda_{3\chi}}{\Lambda_{\chi\chi}}.
\end{split}
\end{align}

By neglecting the electroweak VEVs in the matrix ${M}_{hh}^{2}$, it is obtained that ${M}_{hh}^{2}\approx M_{\mathrm{C}}^{2}$. Thus, ${M}_{hh}^{2}$ should have the two mass eigenvalues $m_{H_{1,2}}^{2}\approx m_{H^{\pm}_{1,2}}^{2}$ at TeV scale and a third one $m_{h}^{2}$ at hundreds of GeV which would be zero if the electroweak vacuum $v$ is neglected. However, the non-vanishing determinant of $M_{hh}^{2}$ shows the existence of the smallest eigenvalue, which can be obtained by dividing the determinant of $M_{hh}^{2}$ by the product of the two largest eigenvalues
\begin{align}
m_{h}^{2} \approx \frac{\mathrm{Det[{M}_{hh}^{2}]}}{m_{H_{1}}^{2}m_{H_{2}}^{2}} \approx \Lambda_{hh} v^2 =
\left(\sum_{i=1}^{i=3} \widetilde{\Lambda}_{ij} v_i^2 v_j^2\right)v^{2}.
\end{align}
where $\Lambda_{hh}$ is the effective coupling constant of the 125 GeV Higgs boson.

The mixing matrix $R_{\mathrm{even}}^{hh}$, which diagonalizes $M_{hh}^{2}$, can be approximated to $R_{\mathrm{C}}$ because the method employed in the eigenvalue search. Thus, the corresponding mixing angles of $R_{\mathrm{even}}^{hh}$ are
\begin{align}
\tan^{2} \theta_{12}^{h} \approx \frac{v_{2}^{2}}{v_{1}^{2}},
\quad \tan^{2} \theta_{13}^{h} \approx \frac{v_{3}^{2}}{v_{1}^{2}+v_{2}^{2}},
\quad \tan^{2} \theta_{23}^{h} \approx 0
\end{align}
Finally, the transformation $R_{\mathrm{even}}^{B}$ which diagonalizes each one of the blocks after the see-saw procedure turns out to be
\begin{align}
R_{\mathrm{even}}^{B} = 
\begin{pmatrix}
	R_{\mathrm{even}}^{hh}	&	0	\\
	0	&	1
\end{pmatrix}.
\end{align}

\subsection*{Summary of masses of the scalar sector}
The scalar sector of the model includes:
\begin{itemize}
\item Three pairs of charged bosons: one pair corresponding to the $W_{\mu}$'s Goldstone bosons $G_{W}^{\pm}$ and two pairs of physical charged scalars with masses given by
\begin{align*}
&m_{H^{\pm}_{1,2}}^{2}\approx 
\frac{{f_{2} } \left(v_1^2+v_2^2\right) {v_{\chi} }}{8 v_1 v_2}+\frac{{f_{3} } \left(v_1^2+v_3^2\right) {v_{\chi} }}{8 v_1 v_3}\\
\pm & \sqrt{\tfrac{f_{2} ^2 \left(v_1^2+v_2^2\right){}^2 v_{\chi}^{2}}{64 v_1^2 v_2^2}-\tfrac{{f_{2} } {f_{3} } \left(v_1^4-v_2^2 v_3^2\right) v_{\chi}^{2}}{32 v_1^{2} v_2 v_3}+\tfrac{f_{3} ^2 \left(v_1^2+v_3^2\right){}^2 v_{\chi}^{2}}{64 v_1^2 v_3^2}}\nonumber .
\end{align*}
\item Four CP-odd bosons: two Goldstone bosons $G_{Z}$ and $G_{Z'}$ corresponding to the gauge fields $Z_{\mu}$ and $Z_{\mu}'$, respectively, and two physical CP-odd scalars with masses given by
\begin{align*}
&m_{A_{1,2}}^{2}\approx 
\frac{{f_{2} } \left(v_1^2+v_2^2\right) {v_{\chi} }}{8 v_1 v_2}+\frac{{f_{3} } \left(v_1^2+v_3^2\right) {v_{\chi} }}{8 v_1 v_3}\\
\pm & \sqrt{\tfrac{f_{2} ^2 \left(v_1^2+v_2^2\right){}^2 v_{\chi}^{2}}{64 v_1^2 v_2^2}-\tfrac{{f_{2} } {f_{3} } \left(v_1^4-v_2^2 v_3^2\right) v_{\chi}^{2}}{32 v_1^{2} v_2 v_3}+\tfrac{f_{3} ^2 \left(v_1^2+v_3^2\right){}^2 v_{\chi}^{2}}{64 v_1^2 v_3^2}}\nonumber .
\end{align*}
\item Four CP-even bosons: the SM-Higgs boson with mass given by $m_{H}^{2} = \Lambda_{hh}v^{2}$,
two new CP-even scalar with masses given by
\begin{align*}
&m_{H_{1,2}}^{2}\approx 
\frac{{f_{2} } \left(v_1^2+v_2^2\right) {v_{\chi} }}{8 v_1 v_2}+\frac{{f_{3} } \left(v_1^2+v_3^2\right) {v_{\chi} }}{8 v_1 v_3}\\
\pm & \sqrt{\tfrac{f_{2} ^2 \left(v_1^2+v_2^2\right){}^2 v_{\chi}^{2}}{64 v_1^2 v_2^2}-\tfrac{{f_{2} } {f_{3} } \left(v_1^4-v_2^2 v_3^2\right) v_{\chi}^{2}}{32 v_1^{2} v_2 v_3}+\tfrac{f_{3} ^2 \left(v_1^2+v_3^2\right){}^2 v_{\chi}^{2}}{64 v_1^2 v_3^2}}\nonumber ,
\end{align*}
and a CP-even boson with mass given by $\lambda_{\chi\chi}v_{\chi}^{2}$.
\end{itemize}

Finally, the scalar sector of the model in the reference \cite{MantillaDic2016} can be recovered by neglecting $v_{3}$ since the previous model has two doublets and one singlet, in contrast with the three doublets and the singlet of this model. 


\section{Fermion masses}
\label{sect:Fermion-masses}

First of all, the fermions of each sector can be described employing two basis: the flavor basis $\mathbf{F}$ or the mass basis $\mathbf{f}$. In the flavor basis, after the Yukawa Lagrangian is evaluated at VEVs, the mass terms can be writing as
\begin{equation}
-\mathcal{L}_{F} = \overline{\mathbf{F}_{L}}\mathbb{M}_{F} \mathbf{F}_{R} + \mathrm{h.c.}
\end{equation}
Since the mass matrix $\mathbb{M}_{F}$ is not Hermitian, it has to be diagonalized by the biunitary transformation
\begin{equation}
\mathbb{M}_{F}^{\mathrm{diag}} = \left(\mathbb{V}^{F}_{L} \right)^{\dagger}\mathbb{M}_{F}\mathbb{V}^{F}_{R},
\end{equation}
and consequently the mass and flavor bases will be related via the mixing matrices $\mathbb{V}^{F}_{L}$ and $\mathbb{V}^{F}_{R}$ in the following way
\begin{equation}
\mathbf{F}_{L} = \mathbb{V}^{F}_{L}\mathbf{f}_{L},	\qquad
\mathbf{F}_{R} = \mathbb{V}^{F}_{R}\mathbf{f}_{R}.
\end{equation}

In particular, the left-handed mixing matrix can be expressed as the product of two mixing matrices
\begin{equation} 
\label{eq:Left-handed-Mixing-Matrix-splitting}
\mathbb{V}^{F}_{L} = \mathbb{V}^{F}_{L,\mathrm{SS}}\mathbb{V}^{F}_{L,\mathrm{B}}.
\end{equation}
The former matrix rotates out the exotic fermions through a see-saw procedure by taking into account the fact that $v_{\chi}\gg v_{1,2,3}$. For this, first we splits the whole symmetric mass matrices in blocks ($\mathbb{M}_{F}\mathbb{M}_{F}^{\dagger}$ for charged fermions and $\mathbb{M}_{N}$ for neutrinos)\cite{grimus2001seesaw}
\begin{equation}
\mathbb{M}_{F}^{\mathrm{sym}} = \begin{pmatrix}
\mathcal{M}^{f}_{3\times 3}	&	\mathcal{M}^{f\mathcal{F}}_{3\times n}	\\
\mathcal{M}^{\mathcal{F}f}_{n\times 3}	&	\mathcal{M}^{\mathcal{F}}_{n\times n}
\end{pmatrix},
\end{equation}
where $\mathcal{M}^{\mathcal{F}f}=\left(\mathcal{M}^{f\mathcal{F}} \right)^{\mathrm{T}}$ and $n$ is the number of exotic fermions for each sector (1 for up-quarks, 2 for down-quarks and charged leptons, and 6 for neutrinos). The see-saw rotation matrix is
\begin{equation}
\mathbb{V}^{F}_{L,\mathrm{SS}} = \begin{pmatrix}
1	&	\Theta^{F\dagger}_{L}	\\	-\Theta^{F}_{L}	&	1
\end{pmatrix},
\end{equation}
where $\Theta^{F}_{L}=\left(\mathcal{M}^{\mathcal{F}}  \right)^{-1}\mathcal{M}^{\mathcal{F}f}$. The resulting block-diagonal mass matrix is 
\begin{equation}
\left( \mathbb{V}^{F}_{L,\mathrm{SS}} \right)^{\mathrm{T}} \mathbb{M}_{F}^{\mathrm{sym}} \mathbb{V}^{F}_{L,\mathrm{SS}} 
 = \begin{pmatrix}
m^{\mathrm{sym}}_{F,\mathrm{SM}}&	0_{3\times n}\\	0_{n\times 3}&	M^{\mathrm{sym}}_{F,\mathrm{exot}}
\end{pmatrix},
\end{equation}
where $m^{\mathrm{sym}}_{F,\mathrm{SM}}$ is the SM mass matrix given by 
\begin{equation}
m^{\mathrm{sym}}_{F,\mathrm{SM}}\approx \mathcal{M}^{f}-\mathcal{M}^{f\mathcal{F}}\left(\mathcal{M}^{\mathcal{F}}  \right)^{-1}\mathcal{M}^{\mathcal{F}f}
\label{eq:see-saw-formula}
\end{equation}
and $M^{\mathrm{sym}}_{F,\mathrm{exot}}\approx \mathcal{M}^{\mathcal{F}}$ is the exotic mass matrix. 
The latter matrix in eq. \eqref{eq:Left-handed-Mixing-Matrix-splitting}, $\mathbb{V}^{F}_{L,\mathrm{B}}$ describes the diagonalization of $m^{\mathrm{sym}}_{F,\mathrm{SM}}$ and $M^{\mathrm{sym}}_{F,\mathrm{exot}}$. It has the structure
\begin{equation}
\mathbb{V}^{F}_{\mathrm{B}} = \begin{pmatrix}
{V}^{F}_{\mathrm{SM}}	&	0_{3\times n}	\\	0_{n\times 3}	&	{V}^{F}_{\mathrm{exot}}
\end{pmatrix}
\end{equation}
where ${V}^{F}_{\mathrm{SM}}$ is parametrized by
\begin{equation}
\label{eq:SM-generic-mixing-matrix}
{V}^{F}_{\mathrm{SM}}=
R_{13}(\theta_{13}^{F},\delta_{13}^{F})
R_{23}(\theta_{23}^{F},\delta_{23}^{F})
R_{12}(\theta_{12}^{F},\delta_{12}^{F})
\end{equation}
and the matrices $R_{ij}$ are
\begin{subequations}
\begin{align}
R_{12}(\theta_{12}^{F}) &= \begin{pmatrix}
c_{12}^{F}&	s_{12}^{F}	&	0\\	-s_{12}^{F*}	&c_{12}^{F}	&0	\\	0&0&1
\end{pmatrix},	\\
R_{13}(\theta_{13}^{F}) &= \begin{pmatrix}
c_{13}^{F}	&0	&s_{13}^{F}	\\0	&1	&0	\\-s_{13}^{F*}	&0	&c_{13}^{F}
\end{pmatrix},	\\
R_{23}(\theta_{23}^{F}) &= \begin{pmatrix}
1&0&0\\	0&	c_{23}^{F}	&s_{23}^{F}	\\0	&-s_{23}^{F*}	&c_{23}^{F}
\end{pmatrix},
\end{align}
\end{subequations}
$c_{ij}^{F}=\cos \theta_{ij}^{F}$ and $s_{ij}^{F}=\sin \theta_{ij}^{F}\exp\left(i\delta_{ij}^{F}\right)$. The angles $\theta_{ij}^{F}$ are specified by their tangents $t_{ij}^{F}=\tan \theta_{ij}^{F}$ which could be calculated exactly or approximately using the vaccum hierarchy of the three Higgs doublet outlined in the section \ref{sect:model} . On the other hand, the Dirac phases $\delta_{ij}^{F}$ can be chosen in such a way that they correspond to the experimental measurements. Regarding to neutrinos, the Majorana phases have to be included (see equation \eqref{eq:PMNS-matrix-parametrization}).

The mass matrices, their mass eigenvalues and mixing angles (involving SM and exotic fermions) can be obtained by using the vacuum hierarchy of the Higgs doublets 
as shown in the next subsections.


\subsection{Up-like quarks}

The up-like quark sector is described in the bases $\mathbf{U}$ and $\mathbf{u}$, where the former is the flavor basis while the latter is the mass basis
\begin{equation}
\begin{split}
\mathbf{U}&=(u^{1},u^{2},u^{3},\mathcal{T}),\\
\mathbf{u}&=(u, c, t, T).
\end{split}
\end{equation}
The mass term in the flavor basis turns out to be
\begin{equation}
-\mathcal{L}_{U} = \overline{\mathbf{U}_{L}} \mathbb{M}_{U} \mathbf{U}_{R} + \mathrm{h.c.},
\end{equation}
where $\mathbb{M}_{U}$ is
\begin{equation}
\mathbb{M}_{U} = \frac{1}{\sqrt{2}}
\begin{pmatrix}
h_{3 u}^{1 1}v_{3}	&	h_{2 u}^{1 2}v_{2}	&h_{3 u}^{1 3}v_{3}	&h_{2 \mathcal{T}}^{1}v_{2}	\\
0	&	h_{1 u}^{2 2}v_{1}	&	0	&h_{1 \mathcal{T}}^{2}v_{1}	\\
h_{1 u}^{3 1}v_{1}	&	0	&	h_{1 u}^{3 3}v_{1}	&	0	\\
0	&	g_{\chi u}^{2}v_{\chi}	&	0	&	g_{\chi \mathcal{T}}v_{\chi}
\end{pmatrix}.
\end{equation}

Since the determinant of $\mathbb{M}_{U}$ is non-vanishing, the four up-like quarks acquire masses. The mass eigenvalues can be calculated by applying different see-saw schemes: the first one rotates out the exotic $\mathcal{T}$ quark, while the second consists on taking advantage of the large hierarchy between $t$ quark and the lightest ones in the VEVs $v_{1}>v_{2}>v_{3}$. Consequently, the four mass eigenvalues are
\begin{equation}
\label{eq:Up-Quarks-masses}
\begin{split}
m_{u}^{2}&=\frac{\left(h_{3 u}^{1 1}h_{1 u}^{3 3}-h_{3 u}^{1 3}h_{1 u}^{3 1}\right)^{2}}{(h_{1 u}^{3 3})^{2}+(h_{1 u}^{3 1})^{2}}
\frac{v_{3}^{2}}{2},	\\
m_{c}^{2}&=\frac{\left(h_{1 u}^{2 2}g_{\chi \mathcal{T}}-h_{1 \mathcal{T}}^{2}g_{\chi u}^{2}\right)^{2}}
{(g_{\chi \mathcal{T}})^{2}+(g_{\chi u}^{2})^{2}}\frac{v_{1}^{2}}{2},	\\
m_{t}^{2}&=\left[(h_{1 u}^{3 3})^{2}+(h_{1 u}^{3 1})^{2} \right]\frac{v_{1}^{2}}{2},\\
m_{T}^{2}& = \left[(g_{\chi \mathcal{T}})^{2}+(g_{\chi u}^{2})^{2} \right]\frac{v_{\chi}^{2}}{2},
\end{split}
\end{equation}
and the corresponding left-handed rotation matrix can be expressed by
\begin{equation}
\mathbb{V}^{U}_{L} = \mathbb{V}^{U}_{L,\mathrm{SS}}\mathbb{V}^{U}_{L,\mathrm{B}}, 
\end{equation}
where the see-saw angle is
\begin{equation}
\Theta^{U\dagger}_{L} = 
\begin{pmatrix}
\frac{h_{2 \mathcal{T}}^{1} g_{\chi \mathcal{T}}+h_{2 u}^{1 2} g_{\chi u}^{2}}{\left(g_{\chi \mathcal{T}} \right)^{2}+\left(g_{\chi u}^{2} \right)^{2}}\frac{v_{2}}{v_{\chi}}	\\
\frac{h_{1 \mathcal{T}}^{2} g_{\chi \mathcal{T}}+h_{1 u}^{2 2} g_{\chi u}^{2}}{\left(g_{\chi \mathcal{T}} \right)^{2}+\left(g_{\chi u}^{2} \right)^{2}}\frac{v_{1}}{v_{\chi}}	\\
0
\end{pmatrix}
\end{equation}
while $\mathbb{V}^{U}_{L,\mathrm{B}}$ diagonalizes only the SM-up quarks. Its angles are given by
\begin{equation}
\begin{split}
t_{12}^{U} &= \frac{h_{2 u}^{1 2} g_{\chi \mathcal{T}}-h_{2 \mathcal{T}}^{1} g_{\chi u}^{2}}{h_{1 u}^{2 2} g_{\chi \mathcal{T}}-h_{1 \mathcal{T}}^{2}g_{\chi u}^{2}} \frac{v_{2}}{v_{1}},\\
t_{13}^{U} &= \frac{h_{3 u}^{1 3} h_{1 u}^{3 3}+h_{3 u}^{1 1} h_{1 u}^{3 1}}{\left(h_{1 u}^{3 3}\right)^{2}+\left(h_{1 u}^{3 1}\right)^{2}}\frac{v_{3}}{v_{1}},\\
t_{23}^{U} &= 0.
\end{split}
\end{equation}

The heavy quarks $T$ and $t$ acquire masses at tree-level through $v_{\chi}$ and $v_{1}$, respectively. The $c$ quark acquire mass also through $v_{1}$, however, this exhibits two suppression mechanisms: by the see-saw with the exotic quark $T$, and the difference of the Yukawa coupling constants. 
Finally, the $u$ quark acquire mass through $v_{3}$ with the same suppression mechanisms of $c$ quark but with $t$ instead of $T$. 

\subsection{Down-like quarks}
The down-like quarks are described in the bases $\mathbf{D}$ and $\mathbf{d}$, where the former is the flavor basis while the latter is the mass basis
\begin{equation}
\begin{split}
\mathbf{D}&=(d^{1},d^{2},d^{3},\mathcal{J}^{1},\mathcal{J}^{2}),\\
\mathbf{d}&=(d, s, b, J^{1}, J^{2}).
\end{split}
\end{equation}
The mass term in the flavor basis is
\begin{equation}
-\mathcal{L}_{D} = \overline{\mathbf{D}_{L}} \mathbb{M}_{D} \mathbf{D}_{R} + \mathrm{h.c.},
\end{equation}
where $\mathbb{M}_{D}$ turns out to be 
\begin{equation}
\label{eq:Down-mass-matrix}
\mathbb{M}_{D} = \frac{1}{\sqrt{2}}
\begin{pmatrix}
0	&	0	&	0	&	h_{1 \mathcal{J}}^{1 1}v_{1}	&	h_{1 \mathcal{J}}^{1 2}v_{1}	\\
h_{3 d}^{2 1}v_{3}	&	h_{3 d}^{2 2}v_{3}	&	h_{3 d}^{2 3}v_{3}	&	h_{2 \mathcal{J}}^{2 1}v_{2}	&	h_{2 \mathcal{J}}^{2 2}v_{2}	\\
h_{2 d}^{3 1}v_{2}	&	h_{2 d}^{3 2}v_{2}	&	h_{2 d}^{3 3}v_{2}	&	h_{3 \mathcal{J}}^{3 1}v_{3}	&	h_{3 \mathcal{J}}^{3 2}v_{3}	\\
0	&	0	&	0	&	g_{\chi \mathcal{J}}^{1}v_{\chi}	&0\\
0	&	0	&	0	&	0	&	g_{\chi \mathcal{J}}^{2}v_{\chi}
\end{pmatrix}
\end{equation}

\begin{figure}
\centering
\includegraphics[scale=1]{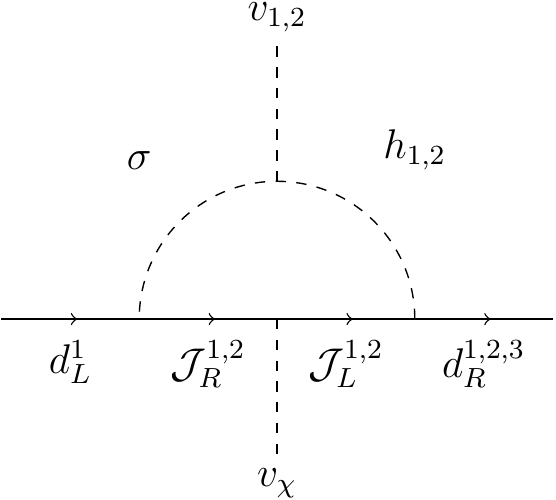}
\caption{1-loop correction to the $d^{1}$-quark propagator.}
\label{fig:One-Loop}
\end{figure}

Unlike the previous cases, the determinant of $\mathbb{M}_{D}$ vanishes. Actually, the rank of the mass matrix is not five but four. Consequently, the lightest quark $d$ remains massless. However, this quark can generate a small mass through radiative corrections according to figure \ref{fig:One-Loop}. The contribution of this diagram is
\begin{equation}
\begin{split}
\Sigma_{d}^{1k} &= \sum_{i=1,2} \frac{f_{\sigma}{g_{\sigma d}^{i 1}}^{*}{h_{k \mathcal{J}}^{k i}}v_{k}}{(4\pi)^{2}m_{Ji}}
C_{0}\left( \frac{m_{\sigma}}{m_{Ji}},\frac{m_{hk}}{m_{Ji}} \right)
\end{split}
\end{equation}
where $k=1,2,3$, $f_{\sigma}$ is the trilinear coupling constant involving $\sigma$ and doublets $\Phi_{1,2,3}$, and the function $C_{0}\left( x,y \right)$ is given by\cite{hernandez2013radiative}
\begin{equation}
\begin{split}
C_{0}\left( x,y \right) &= \frac{1}{(1-x^{2})(1-y^{2})(x^{2}-y^{2})}	\\
	&\times \left\lbrace x^{2}y^{2}\ln\left(\frac{x^{2}}{y^{2}} \right)
	- x^{2}\ln x^{2} + y^{2}\ln y^{2} \right\rbrace 
\end{split}
\end{equation}
Thus, up to one-loop correction, the mass matrix is
\begin{equation}
\mathbb{M}_{D} = \frac{1}{\sqrt{2}}
\begin{pmatrix}
\Sigma_{d}^{11}	&	\Sigma_{d}^{12}	&	\Sigma_{d}^{13}	&	h_{1 \mathcal{J}}^{1 1}v_{1}	&	h_{1 \mathcal{J}}^{1 2}v_{1}	\\
h_{3 d}^{2 1}v_{3}	&	h_{3 d}^{2 2}v_{3}	&	h_{3 d}^{2 3}v_{3}	&	h_{2 \mathcal{J}}^{2 1}v_{2}	&	h_{2 \mathcal{J}}^{2 2}v_{2}	\\
h_{2 d}^{3 1}v_{2}	&	h_{2 d}^{3 2}v_{2}	&	h_{2 d}^{3 3}v_{2}	&	h_{3 \mathcal{J}}^{3 1}v_{3}	&	h_{3 \mathcal{J}}^{3 2}v_{3}	\\
0	&	0	&	0	&	g_{\chi \mathcal{J}}^{1}v_{\chi}	&0\\
0	&	0	&	0	&	0	&	g_{\chi \mathcal{J}}^{2}v_{\chi}
\end{pmatrix}	
\end{equation}
whose determinant does not vanish. Its diagonalization is straightforward by considering the hierarchy $\Sigma_{dj}\ll v_{3} \ll v_{2} \ll v_{\chi}$. 

The masses of the $d$ and $s$ quarks are given by
\begin{widetext}
\begin{equation}
\label{eq:Down-Quarks-masses-Light}
\begin{split}
m_{d}^{2} &= \frac{\left[
\left(\Sigma_{d}^{11}h_{3 d}^{2 2}-\Sigma_{d}^{12}h_{3 d}^{2 1} \right)h_{2 d}^{3 3} + 
\left(\Sigma_{d}^{13}h_{3 d}^{2 1}-\Sigma_{d}^{11}h_{3 d}^{2 3} \right)h_{2 d}^{3 2} + 
\left(\Sigma_{d}^{12}h_{3 d}^{2 3}-\Sigma_{d}^{13}h_{3 d}^{2 2} \right)h_{2 d}^{3 1}
\right]^2}{
\left[(h_{3 d}^{2 1})^2+(h_{3 d}^{2 2})^2 \right](h_{2 d}^{3 3})^2 + 
\left[(h_{3 d}^{2 3})^2+(h_{3 d}^{2 1})^2 \right](h_{2 d}^{3 2})^2 + 
\left[(h_{3 d}^{2 2})^2+(h_{3 d}^{2 3})^2 \right](h_{2 d}^{3 1})^2},	\\
m_{s}^{2} &= \frac{
\left[(h_{3 d}^{2 1})^2+(h_{3 d}^{2 2})^2 \right](h_{2 d}^{3 3})^2 + 
\left[(h_{3 d}^{2 3})^2+(h_{3 d}^{2 1})^2 \right](h_{2 d}^{3 2})^2 + 
\left[(h_{3 d}^{2 2})^2+(h_{3 d}^{2 3})^2 \right](h_{2 d}^{3 1})^2}
{(h_{2 d}^{3 3})^2+(h_{2 d}^{3 2})^2+(h_{2 d}^{3 1})^2}\frac{v_{3}^{2}}{2},
\end{split}
\end{equation}
\end{widetext}
while the masses of the heaviest $b$, $J^{1}$ and $J^{2}$ are
\begin{equation}
\label{eq:Down-Quarks-masses-Heavy}
\begin{split}
m_{b}^{2} &= \left[(h_{2 d}^{3 3})^2+(h_{2 d}^{3 2})^2+(h_{2 d}^{3 1})^2 \right]\frac{v_{2}^{2}}{2},	\\
m_{J1}^{2} &= (g_{\chi \mathcal{J}}^{1})^2\frac{v_{\chi}^{2}}{2},	\quad 
m_{J2}^{2} = (g_{\chi \mathcal{J}}^{2})^2\frac{v_{\chi}^{2}}{2}.
\end{split}
\end{equation}

The corresponding left-handed rotation matrix is
\begin{equation}
\mathbb{V}^{D}_{L} = \mathbb{V}^{D}_{L,\mathrm{SS}}\mathbb{V}^{D}_{L,\mathrm{B}}, 
\end{equation}
where the see-saw angle which rotates out species $J^{i}$ is
\begin{equation}
\Theta^{D\dagger}_{L} = 
\begin{pmatrix}
\dfrac{h_{1 \mathcal{J}}^{1 1}}{g_{\chi \mathcal{J}}^{1}}\dfrac{v_{1}}{v_{\chi}}	&	
\dfrac{h_{1 \mathcal{J}}^{1 2}}{g_{\chi \mathcal{J}}^{1}}\dfrac{v_{2}}{v_{\chi}}	\\
\dfrac{h_{2 \mathcal{J}}^{1 1}}{g_{\chi \mathcal{J}}^{1}}\dfrac{v_{2}}{v_{\chi}}	&	
\dfrac{h_{2 \mathcal{J}}^{1 2}}{g_{\chi \mathcal{J}}^{2}}\dfrac{v_{2}}{v_{\chi}}	\\
\dfrac{h_{3 \mathcal{J}}^{1 1}}{g_{\chi \mathcal{J}}^{1}}\dfrac{v_{3}}{v_{\chi}}	&	
\dfrac{h_{3 \mathcal{J}}^{1 2}}{g_{\chi \mathcal{J}}^{2}}\dfrac{v_{3}}{v_{\chi}}
\end{pmatrix},
\end{equation}
and the SM angles of $\mathbb{V}^{D}_{L,\mathrm{B}}$ are given by
\begin{equation}
\begin{split}
t_{12} &= \frac{{\Sigma}_{d}^{11}h_{3 d}^{2 1}+\Sigma_{d}^{12}h_{3 d}^{2 2}+\Sigma_{d}^{13}h_{3 d}^{2 3}}
{(h_{3 d}^{2 1})^2+(h_{3 d}^{2 2})^2+(h_{3 d}^{2 3})^2v_{3}},	\\
t_{13} &= \frac{\Sigma_{d}^{11}h_{2 d}^{3 1}+\Sigma_{d}^{12}h_{2 d}^{3 2}+\Sigma_{d}^{13}h_{2 d}^{3 3}}
{(h_{2 d}^{3 1})^2+(h_{2 d}^{3 2})^2+(h_{2 d}^{3 3})^2v_{2}},	\\
t_{23} &= \frac{h_{3 d}^{2 1}h_{2 d}^{3 1}+h_{3 d}^{2 2}h_{2 d}^{3 2}+h_{3 d}^{2 3}h_{2 d}^{3 3}}
{(h_{2 d}^{3 1})^2+(h_{2 d}^{3 2})^2+(h_{2 d}^{3 3})^2}\frac{v_{3}}{v_{2}}.
\end{split}
\end{equation}

The heaviest quarks $J^{1}$ and $J^{2}$ acquire masses at TeV scale due to $v_{\chi}$, while the $b$ quark obtaine mass through $v_{2}$ at GeV. The strange quark acquire mass proportional to $v_{3}$ at hundreds of MeV with the suppression due to the $b$ quark. The lightest $d$ quark did not acquire mass at tree-level but at one-loop, where tha radiative correction works as a suppression mechanism.

As an alternative scenario, if $\sigma$ acquires a VEV $v_{\sigma}$ smaller than $v_{3}$, the entries of the fourth and fifth rows of the matrix in eq. \eqref{eq:Down-mass-matrix} are not null. In this case, the mass matrix is
\begin{equation}
\label{eq:Down-mass-matrix-vSigma}
\mathbb{M}_{D} = \frac{1}{\sqrt{2}}
\begin{pmatrix}
0	&	0	&	0	&	h_{1 \mathcal{J}}^{1 1}v_{1}	&	h_{1 \mathcal{J}}^{1 2}v_{1}	\\
h_{3 d}^{2 1}v_{3}	&	h_{3 d}^{2 2}v_{3}	&	h_{3 d}^{2 3}v_{3}	&	h_{2 \mathcal{J}}^{2 1}v_{2}	&	h_{2 \mathcal{J}}^{2 2}v_{2}	\\
h_{2 d}^{3 1}v_{2}	&	h_{2 d}^{3 2}v_{2}	&	h_{2 d}^{3 3}v_{2}	&	h_{3 \mathcal{J}}^{3 1}v_{3}	&	h_{3 \mathcal{J}}^{3 2}v_{3}	\\
g_{\sigma d}^{1 1}v_{\sigma}	&	g_{\sigma d}^{1 2}v_{\sigma}	&	g_{\sigma d}^{1 3}v_{\sigma}	&	g_{\chi \mathcal{J}}^{1}v_{\chi}	&0\\
g_{\sigma d}^{2 1}v_{\sigma}	&	g_{\sigma d}^{2 2}v_{\sigma}	&	g_{\sigma d}^{2 3}v_{\sigma}	&	0	&	g_{\chi \mathcal{J}}^{2}v_{\chi}
\end{pmatrix}	
\end{equation}
whose determinant is non-vanishing and consequently the $d$ quark is massive, with mass given by
\begin{equation}
m_{d}^{2} = \frac{(g_{\sigma d}^{1 1})^{2}(h_{1 \mathcal{J}}^{1 2})^{2}}{(g_{\chi \mathcal{J}}^{1}v_{\chi})^{2}}\frac{v_{\sigma}^{2}v_{1}^{2}}{v_{\chi}^{2}}
\end{equation}
where for simplicity $h_{3 d}^{2 1}$, $h_{3 d}^{2 3}$ and $h_{2 d}^{3 1}$ has been set to zero in order to simplify the expression, 

\subsection{Neutral leptons}
The neutrinos involve both Dirac and Majorana masses in their Yukawa Lagrangian. 
The flavor and mass basis are respectively
\begin{equation}
\begin{split}
\mathbf{N}_{L}&=(\nu^{e,\mu,\tau}_{L},{\nu^{e,\mu,\tau}_{R}}^{C},{\mathcal{N}^{e,\mu,\tau}_{R}}^{C}),\\
\mathbf{n}_{L}&=(\nu^{1,2,3}_{L},N^{1,2,3}_{L},\tilde{N}^{1,2,3}_{L}).
\end{split}
\end{equation}
The mass term expressed in the flavor basis is
\begin{equation}
-\mathcal{L}_{N} = \frac{1}{2} \overline{\mathbf{N}_{L}^{C}} \mathbb{M}_{N} \mathbf{N}_{L},
\end{equation}
where the mass matrix has the following block structure
\begin{equation}
\mathbb{M}_{N} = 
\left(\begin{array}{c c c}
0	&	\mathcal{M}_{\nu}^{\mathrm{T}}	&	0	\\
\mathcal{M}_{\nu}	&	0	&	\mathcal{M}_{\mathcal{N}}^{\mathrm{T}}	\\
0	&	\mathcal{M}_{\mathcal{N}}	&	M_{\mathcal{N}}
\end{array}\right),
\end{equation}
with $\mathcal{M}_{\mathcal{N}} = \mathrm{diag}\left( \begin{matrix}
h_{\mathcal{N}}^{1},h_{\mathcal{N}}^{2},h_{\mathcal{N}}^{3}
\end{matrix} \right)\frac{v_{\chi}}{\sqrt{2}}$ the Dirac mass in the ($\nu_{R}^{C}$, $\mathcal{N}_{R}$) basis, and
\begin{equation}
\label{eq:m_nu_original_parameters}
\mathcal{M}_{\nu} = \frac{v_{3}}{\sqrt{2}}\left(\begin{matrix}
h_{3 \nu}^{e e}	&	h_{3 \nu}^{e \mu}	&	h_{3 \nu}^{e \tau}	\\
h_{3 \nu}^{\mu e}&	h_{3 \nu}^{\mu \mu}	&	h_{3 \nu}^{\mu \tau}	\\
0	&	0	&	0	\end{matrix}\right),
\end{equation}
is a Dirac mass matrix for ($\nu_{L}$, $\nu_{R}$). $M_{\mathcal{N}} = \mu_{\mathcal{N}} \mathbb{I}_{3\times 3}$ is the Majorana mass of $\mathcal{N}_{R}$.

By employing the inverse SSM, taking into account the hierarchy $v_{\chi}\gg v_{3}\gg |M_{\mathcal{N}}|$, we find that 
\begin{equation}
\left(\mathbb{V}_{L,\mathrm{SS}}^{N}\right)^{\dagger}\mathbb{M}_{N} \mathbb{V}_{L,\mathrm{SS}}^{N} = 
\begin{pmatrix}
m_{\nu}	&	0	&	0	\\
0	&	m_{N}	&	0	\\
0	&	0	&	m_{\tilde{N}}
\end{pmatrix}
\end{equation}
where the resultant $3\times 3$ blocks are\cite{catano2012neutrino,dias2012simple}
\begin{equation}
\label{eq:Neutrino-block-mass-matrices}
\begin{split}
m_{\nu} &=	\mathcal{M}_{\nu}^{\mathrm{T}} \left( \mathcal{M}_{\mathcal{N}} \right)^{-1} M_{\mathcal{N}} \left( \mathcal{M}_{\mathcal{N}}^{\mathrm{T}} \right)^{-1} \mathcal{M}_{\nu},	\\
M_{N} &\approx \mathcal{M}_{\mathcal{N}}-{M}_{\mathcal{N}},	\quad	
M_{\tilde{N}}\approx \mathcal{M}_{\mathcal{N}}+{M}_{\mathcal{N}}.
\end{split}
\end{equation}
The most important details of these matrices are discussed in section \ref{sect:PMNS-matrix}. 

\subsection{Charged leptons}
The charged leptons are described in the bases $\mathbf{E}$ and $\mathbf{e}$, where the former is the flavor basis while the latter is the mass basis
\begin{equation}
\begin{split}
\mathbf{E}&=(e^{e},e^{\mu} , e^{\tau}, \mathcal{E}^{1}, \mathcal{E}^{2}),\\
\mathbf{e}&=(e, \mu, \tau, E^{1}, E^{2}).
\end{split}
\end{equation}
The mass term obtained from the Yukawa Lagrangian is
\begin{equation}
-\mathcal{L}_{E} = \overline{\mathbf{E}_{L}}\mathbb{M}_{E} \mathbf{E}_{R} + \mathrm{h.c.}
\end{equation}
where $\mathbb{M}_{E}$ turns out to be 
\begin{equation}
\label{eq:Electron-mass-matrix}
\begin{split}
\mathbb{M}_{E} =  \frac{1}{\sqrt{2}}
\begin{pmatrix}
0	&	h_{3 e}^{e \mu}v_{3}	&	0	&	h_{1 \mathcal{E}}^{e 1}v_{1}	&	0	\\
0	&	h_{3 e}^{\mu \mu}v_{3}	&	0	&	h_{1 \mathcal{E}}^{\mu 1}v_{1}	&	0	\\
h_{3 e}^{\tau e}v_{3}	&	0	&	h_{2 e}^{\tau \tau}v_{2}	&	0	&	0	\\
g_{\chi e}^{1 e}v_{\chi}	&	0	&	0	&	g_{\chi \mathcal{E}}^{1}v_{\chi}	&0\\
0	&	g_{\chi e}^{2 \mu}v_{\chi}	&	0	&0	&	g_{\chi \mathcal{E}}^{2}v_{\chi}
\end{pmatrix}
\end{split}
\end{equation}

The determinant of $\mathbb{M}_{E}$ is non-vanishing ensuring that the five charged leptons acquire masses. Although its eigenvalues and the mixing matrix $\mathbb{V}_{L}^{E}$ have large analytical solutions, we can obtain predictable expressions by implementing the vacuum hierarchy of the Higgs doublets. The resulting eigenvalues are
\begin{equation}
\label{eq:Charged-Lepton-masses}
\begin{split}
m_{e}^{2} &= \frac{\left(h_{3 e}^{e \mu}h_{1 \mathcal{E}}^{\mu 1}-h_{3 e}^{\mu \mu}h_{1 \mathcal{E}}^{e 1}\right)^2}
{(h_{1 \mathcal{E}}^{e 1})^2+(h_{1 \mathcal{E}}^{\mu 1})^2}\frac{v_{3}^{2}}{2},		\\
m_{\mu}^{2} &= \frac{\left(h_{3 e}^{e \mu}h_{1 \mathcal{E}}^{e 1}+h_{3 e}^{\mu \mu}h_{1 \mathcal{E}}^{\mu 1}\right)^2}
{(h_{1 \mathcal{E}}^{e 1})^2+(h_{1 \mathcal{E}}^{\mu 1})^2}\frac{v_{3}^{2}}{2}
+\frac{\left(h_{3 e}^{\tau e}\right)^{2} v_{3}^{2}}{2},	\\
m_{\tau}^{2} &= \frac{\left(h_{2 e}^{\tau \tau}\right)^{2}v_{2}^{2}}{2},	\\
m_{E{1}}^{2} &= \left[\left(g_{\chi \mathcal{E}}^{1}\right)^{2}+\left(g_{\chi e}^{1 e}\right)^{2}\right]\frac{v_{\chi}^{2}}{2},	\\
m_{E{2}}^{2} &= \left[\left(g_{\chi \mathcal{E}}^{2}\right)^2+\left(g_{\chi e}^{2\mu}\right)^2\right]\frac{v_{\chi}^{2}}{2}.
\end{split}
\end{equation}
The exotic charged leptons $E^{1}$ and $E^{2}$ have acquired masses at the TeV scale, while the heaviest SM lepton $\tau$ acquired mass at  the GeV scale, proportional to $v_{2}$. On the other hand, the charged leptons $\mu$ and $e$ have acquired mass through $v_{3}$ which constitutes the smallest VEV. Both of them participate in a sort of see-saw with the matrix entries proportional to $v_{1}$ which help us to suppress their masses. Moreover, the $e$ mass is further suppressed because of the difference between the Yukawa coupling constants in the first equation of (\ref{eq:Charged-Lepton-masses}), which can be assumed to be at the same order of magnitude. 

\subsubsection{Left-handed rotation}
The unitary transformation which diagonalizes the matrix $\mathbb{M}^{E}_{L}=\mathbb{M}_{E}\mathbb{M}_{E}^{\dagger}$ can be split as follows
\begin{equation}
\mathbb{V}^{E}_{L} = \mathbb{V}^{E}_{L,\mathrm{SS}}\mathbb{V}^{E}_{L,\mathrm{B}}.
\end{equation}
The see-saw procedure is done by $\mathbb{V}^{E}_{L,\mathrm{SS}}$ because $\mathbb{M}^{E}_{L}$ has the suited hierarchy in its sub-blocks. The corresponding see-saw angle turns out to be
\begin{equation}
\Theta^{E\dagger}_{L} =\begin{pmatrix}
\frac{h_{1 \mathcal{E}}^{e 1}g_{\chi \mathcal{E}}^{1}v_{1}v_{\chi}}{2 m_{E^{1}}^{2}} &
\frac{h_{3 e}^{e \mu}g_{\chi e}^{2 \mu}v_{3}v_{\chi}}{2 m_{E^{2}}^{2}}\\
\frac{h_{1 \mathcal{E}}^{\mu 1}g_{\chi \mathcal{E}}^{1} v_{1}v_{\chi}}{2 m_{E^{1}}^{2}} &
\frac{h_{3 e}^{\mu \mu}g_{\chi e}^{2 \mu} v_{3}v_{\chi}}{2 m_{E^{2}}^{2}}	\\
\frac{\left( h_{2 e}^{\tau e}g_{\chi e}^{1 e}+h_{2 e}^{\tau \tau}g_{\chi e}^{1 \tau}\right) v_{3}v_{\chi}}{2 m_{E^{1}}^{2}} & 0
\end{pmatrix}. 
\end{equation}
Then, the transformation $\mathbb{V}^{E}_{L,\mathrm{B}}$ only diagonalizes SM leptons with angles given by
\begin{equation}
\label{eq:Electron-SM-Rotation-angles}
\begin{split}
t_{L,12}^{E} &\approx \frac{h_{1E}^{e 1}}{h_{1E}^{\mu 1}} ,\\
t_{L,23}^{E} &\approx -\frac{2 g_{\chi E}^1{}^3 h_{2e}^{\tau e} h_{2e}^{\tau \tau }{}^2}{g_{\chi e}^{e 1}{}^3 h_{1E}^{e 1}{}^2 h_{1E}^{\mu 1}}\frac{v_2{}^2 v_3 }{v_1{}^3 },	\\
t_{L,13}^{E} &\approx \frac{g_{\chi E}^1 h_{2e}^{\tau e}}{g_{\chi e}^{e 1} h_{1E}^{e 1}}\frac{v_{3}}{v_{1}}.
\end{split}
\end{equation}

\subsubsection{Right-handed rotation}
On the other hand, the right-handed matrix $\mathbb{M}^{E}_{R}=\mathbb{M}_{E}^{\dagger}\mathbb{M}_{E}$ cannot be diagonalized by means of the see-saw procedure because the presence of $v_{\chi}^{2}$ terms in the top-left $3\times 3$ block. Therefore, finite angles are required to rotate out any $v_{\chi}^{2}$ in contrast with the diagonalization procedure applied on $\mathbb{M}^{E}_{L}$. These angles can be approximately obtained by neglecting any electroweak vacuum in $\mathbb{M}^{E}_{R}$
\begin{equation}
\mathbb{M}^{E}_{R}\approx
\small{
\frac{v_{\chi }^2}{2}\begin{pmatrix}
 g_{{\chi e}}^{{e1}}{}^2  & 0 & g_{{\chi e}}^{{e1}} g_{{\chi e}}^{{\tau 1}}  & g_{{\chi \mathcal{E} }}^1 g_{{\chi e}}^{{e1}}  & 0 \\
 0 & g_{{\chi e}}^{{\mu 2}}{}^2  & 0 & 0 & g_{{\chi \mathcal{E} }}^2 g_{{\chi e}}^{{\mu 2}}  \\
 g_{{\chi e}}^{{e1}} g_{{\chi e}}^{{\tau 1}}  & 0 & g_{{\chi e}}^{{\tau 1}}{}^2  & g_{{\chi \mathcal{E} }}^1 g_{{\chi e}}^{{\tau 1}}  & 0 \\
 g_{{\chi \mathcal{E} }}^1 g_{{\chi e}}^{{e1}}  & 0 & g_{{\chi \mathcal{E} }}^1 g_{{\chi e}}^{{\tau 1}}  & g_{{\chi \mathcal{E} }}^1{}^2  & 0 \\
 0 & g_{{\chi \mathcal{E} }}^2 g_{{\chi e}}^{{\mu 2}}  & 0 & 0 & g_{{\chi \mathcal{E} }}^2{}^2 	
\end{pmatrix}}
\end{equation}
and diagonalizing it in such a way the masses of the exotic species $E^{1}$ and $E^{2}$ result in the bottom-right block. This rotation may be expressed by the parametrization
\begin{equation}
\mathbb{V}^{E}_{R,v_{\chi}}=
R_{25}(\theta_{R,25}^{E})	R_{34}(\theta_{R,34}^{E})	R_{14}(\theta_{R,14}^{E})
\end{equation}
and the corresponding angles are given by
\begin{equation}
\label{eq:Electron-Right-vX-Rotation-angles}
\begin{split}
t_{R,25}^{E} &\approx \frac{g^{\mu  2}_{\chi e}}{g^{2}_{\chi E}},	\\
t_{R,34}^{E} &\approx \frac{g^{\tau 1}_{\chi e}}{g^{1}_{\chi E}},	\\ 
t_{R,14}^{E} &\approx \frac{g^{e 1}_{\chi e}}
{\sqrt{(g^{  1   }_{\chi \mathcal{E}})^{2}+(g^{\tau 1}_{\chi e})^{2}}}	.
\end{split}
\end{equation}
Consequently, the diagonalization is done by the transformation
\begin{equation}
\mathbb{V}^{E}_{R} = \mathbb{V}^{E}_{R,v_{\chi}}\mathbb{V}^{E}_{R,\mathrm{B}},
\end{equation}
After rotating out $v_{\chi}$ from the top-left $3\times3$ block, it is viable to implement a similar rotation to $\mathbb{V}^{E}_{L,\mathrm{SM}}$ with the following angles
\begin{equation}
\label{eq:Electron-Right-SM-Rotation-angles}
\begin{split}
t_{R,12}^{E} &\approx -\frac{g_{\chi e}^{e 1} \left(h_{1E}^{e 1}{}^2+h_{1E}^{\mu 1}{}^2\right)}{g_{\chi E}^1 (h_{1E}^{e 1} h_{3e}^{e \mu}+h_{1E}^{\mu 1} h_{3e}^{\mu \mu })}\frac{v_{1}}{v_{3}},\\
t_{R,23}^{E} &\approx \frac{g_{\chi E}^1 h_{2e}^{\tau e} (h_{1E}^{e 1} h_{3e}^{e \mu}+h_{1E}^{\mu 1} h_{3e}^{\mu \mu })}{g_{\chi e}^{e 1} h_{2e}^{\tau \tau } \left(h_{1E}^{e 1}{}^2+h_{1E}^{\mu 1}{}^2\right)},\frac{v_3{}^2 }{v_1 v_2 }\\
t_{R,13}^{E} &\approx \frac{g_{\chi E}^1{}^2 h_{2e}^{\tau e} h_{2e}^{\tau \tau }}{g_{\chi e}^{e 1}{}^2 \left(h_{1E}^{e 1}{}^2+h_{1E}^{\mu 1}{}^2\right)}\frac{v_2 v_3 }{v_1{}^2 }.
\end{split}
\end{equation}

Summarizing, the fermion mass hierarchy is induced by the generation of a hierarchy of the vaccum of the Higgs doublets together with the mass matrices obtained from the Yukawa Lagrangian, whose terms are constrained by the non-universal $\mathrm{U(1)}_{X}$ gauge and $\mathbb{Z}_{2}$ discrete symmetries. The fermion masses are outlined in the table \ref{tab:Fermion-masses}.

\vspace{3mm}

\section{Neutrino parameters}
\label{sect:PMNS-matrix}
The consistency of this model with the current neutrino oscillation data shown in the table \ref{tab:Neutrino-data} is tested by exploring the parameter space of the neutral sector of the Yukawa Lagrangian. For simplicity, we choose a basis for $\nu_{R}$ where $\mathcal{M}_{\mathcal{N}}$ is diagonal, and $M_{\mathcal{N}}$ is proportional to the identity
\begin{equation}
\label{eq:R-Neutrino-mass-matrix}
\mathcal{M}_{\mathcal{N}} = \mathrm{diag}\left( \begin{matrix}
h_{\mathcal{N}}^{1},h_{\mathcal{N}}^{2},h_{\mathcal{N}}^{3}
\end{matrix} \right)\frac{v_{\chi}}{\sqrt{2}}
\end{equation}
\begin{equation}
{M}_{\mathcal{N}} = \mu_{\mathcal{N}} \mathbb{I}_{3\times 3}.
\end{equation}
where $\mu_{\mathcal{N}}$ fixes the Majorana mass such that the light neutrinos acquire masses at eV scale. On the other hand, the coupling constants $h_{\mathcal{N}}^{1}$, $h_{\mathcal{N}}^{2}$ and $h_{\mathcal{N}}^{3}$ determine the masses of the heaviest neutrinos. 

By replacing the Dirac mass matrix from \eqref{eq:m_nu_original_parameters} into the light mass eigenvalues in \eqref{eq:Neutrino-block-mass-matrices} the explicit expression of the SM neutrino mass matrix is obtained
\begin{widetext}
\begin{equation}
\label{eq:Neutrino-mass-matrix}
m_{\nu} = \frac{\mu_{\mathcal{N}} v_{3}^{2}}{{\left(h_{\mathcal{N}}^{1}\right)}^{2}v_{\chi}^{2}}
\left( \begin{matrix}
	\left( h_{3 \nu}^{e e}\right)^{2} + \left( h_{3 \nu}^{\mu e} \right)^{2} \rho^{2} 	&
	{h_{3 \nu}^{e e}}\,{h_{3 \nu}^{e \mu}} + {h_{3 \nu}^{\mu e}}\,{h_{3 \nu}^{\mu \mu}}\rho^2 	&
	{h_{3 \nu}^{e e}}\,{h_{3 \nu}^{e \tau}}+ {h_{3 \nu}^{\mu e}}\,{h_{3 \nu}^{\mu \tau}}\rho^2 	\\
	{h_{3 \nu}^{e e}}\,{h_{3 \nu}^{e \mu}} + {h_{3 \nu}^{\mu e}}\,{h_{3 \nu}^{\mu \mu}}\rho^2	&	
	\left( h_{3 \nu}^{e \mu} \right)^{2} + ( h_{3\mu}^{\nu\mu} )^{2} \rho^{2}	&	
	{h_{3 \nu}^{e \mu}}\,{h_{3 \nu}^{e \tau}}+ {h_{3 \nu}^{\mu \mu}}\,{h_{3 \nu}^{\mu \tau}}\rho^2	\\
	{h_{3 \nu}^{e e}}  \,{h_{3 \nu}^{e \tau}}+ {h_{3 \nu}^{\mu e}}  \,{h_{3 \nu}^{\mu \tau}}\rho^2	&	
	{h_{3 \nu}^{e \mu}}\,{h_{3 \nu}^{e \tau}}+ {h_{3 \nu}^{\mu \mu}}\,{h_{3 \nu}^{\mu \tau}}\rho^2	&	
	\left( h_{3 \nu}^{e \tau} \right)^{2} + \left( h_{3 \nu}^{\mu \tau} \right)^{2} \rho^{2}
\end{matrix} \right).
\end{equation}
\end{widetext}
The ratio $\rho={h_{\mathcal{N}}^{1}}/{h_{\mathcal{N}}^{2}}$ describes the heavy neutrino hierarchy. Since the matrix $m_{\nu}$ has null determinant, at least one neutrino is massless. The above matrix is diagonalized by
\begin{equation*}
\left( {V}^{N}_{L,\mathrm{SM}} \right)^{\dagger} m_{\nu} {V}^{N}_{L,\mathrm{SM}} = m_{\nu}^{\mathrm{diag}},
\end{equation*}
which together with ${V}^{E}_{L,\mathrm{SM}}$ constitute the Pontecorvo-Maki-Nakagawa-Sakata (PMNS) matrix \cite{pontecorvo1958inverse,maki1962remarks}
\begin{equation}
\begin{split}
{U}_{\ell} &= \left( {V}^{E}_{L,\mathrm{SM}} \right)^{\dagger} {V}^{N}_{L,\mathrm{SM}}.
\end{split}
\end{equation}
	
The parametrization for the PMNS matrix follows the convention shown in eq. \eqref{eq:SM-generic-mixing-matrix} given by
\begin{equation}
\label{eq:PMNS-matrix-parametrization}
U_{\ell} = D(1,\delta_{2}^{N},\delta_{3}^{N})R_{23}(\theta_{23}^{N})
R_{13}(\theta_{13}^{N},\delta_{13}^{N})R_{12}(\theta_{12}^{N}),
\end{equation}
where $D(1,\delta_{2}^{N},\delta_{3}^{N})$ is the Majorana phase matrix
\begin{equation*}
D(1,\delta_{2}^{N},\delta_{3}^{N})=
\begin{pmatrix}
1&0&0\\0&e^{i\delta_{2}^{N}}&0\\0&0&e^{i\delta_{3}^{N}}
\end{pmatrix}.
\end{equation*}
The angles can be obtained following the convention presented in the PDG \cite{patrignani2016review}
\begin{equation}
\begin{split}
s_{13}^{2} &= \left| U_{e3} \right|^{2},	\\
s_{23}^{2} &= \frac{\left| U_{\mu3} \right|^{2}}{1-\left| U_{e3} \right|^{2}},	\\
s_{12}^{2} &= \frac{\left| U_{  e2  } \right|^{2}}{1-\left| U_{e3} \right|^{2}}.
\end{split}
\label{PMNS-mixing angles}
\end{equation}

The resulting angles obtained from the experimental data are shown in table \ref{tab:Neutrino-data} which have been fitted in the references \cite{esteban2016updated,nufit} where the convention \eqref{PMNS-mixing angles} was employed. 

\begin{table}
\centering
\begin{tabular}{c||cc}
	&	NO	&	IO	\\	\hline\hline
$\sin^{2}\theta_{12}$	&	$0.308^{+0.013}_{-0.012}$	&	$0.308^{+0.013}_{-0.012}$	\\	
$\sin^{2}\theta_{23}$	&	$0.440^{+0.023}_{-0.019}$	&	$0.584^{+0.018}_{-0.022}$	\\	
$\sin^{2}\theta_{13}$	&	$0.02163^{+0.00074}_{-0.00074}$	&	$0.02175^{+0.00075}_{-0.00074}$	\\	
$\delta_{\mathrm{CP}}$	&	$289^{+38}_{-51}$	&	$269^{+39}_{-45}$	\\ 
$\dfrac{\Delta m_{21}^{2}}  {10^{-5}\mathrm{\,eV^{2}}}$	&	$7.49^{+0.19}_{-0.17}$	&	$7.49^{+0.19}_{-0.17}$	\\ 
$\dfrac{\Delta m_{3\ell}^{2}}  {10^{-3}\mathrm{\,eV^{2}}}$	&	$+2.526^{+0.039}_{-0.037}$	&	$-2.518^{+0.038}_{-0.037}$
\end{tabular}
\caption{Three-flavor oscillation parameters fitting  at 1$\sigma$ reported by \cite{esteban2016updated,nufit}. $\ell=1$ for normal ordering (NO) and $2$ for inverted ordering (IO).}
\label{tab:Neutrino-data}
\end{table}

\subsection{Numerical exploration of $m_{\nu}$ consistent with current data}
Since the components of the neutrino mass matrix $m_{\nu}$ are quadratic forms of the Yukawa couplings, it is useful to do some coordinate transformation to simplify them. The Yukawa couplings are expressed in their "Cartesian" fashion, but their "polar" form can be written as
\begin{equation}
\begin{split}
\left\lbrace\begin{split}
     h_{3 \nu}^{e e} &= h_{\nu}^{e}\cos \theta_{\nu}^{e}	\\
\rho h_{3 \nu}^{\mu e} &= h_{\nu}^{e}\sin \theta_{\nu}^{e}
\end{split}\right.,	\\
\left\lbrace\begin{split}
     h_{3 \nu}^{e \mu} &= h_{\nu}^{\mu}\cos \theta_{\nu}^{\mu}	\\
\rho h_{3 \nu}^{\mu \mu} &= h_{\nu}^{\mu}\sin \theta_{\nu}^{\mu}
\end{split}\right.,	\\
\left\lbrace\begin{split}
     h_{3 \nu}^{e \tau} &= h_{\nu}^{\tau}\cos \theta_{\nu}^{\tau}	\\
\rho h_{3 \nu}^{\mu \tau} &= h_{\nu}^{\tau}\sin \theta_{\nu}^{\tau}
\end{split}\right..
\end{split}
\end{equation}

The Dirac mass matrix becomes
\begin{equation}
\label{eq:m_nu_polar_parameters}
\mathcal{M}_{\nu} = \frac{v_{3}}{\sqrt{2}{\rho}}\left(\begin{matrix}
{\rho}h_{\nu}^{e}c_{\nu}^{e}&	{\rho}h_{\nu}^{\mu}c_{\nu}^{\mu}	&	{\rho}h_{\nu}^{\tau}c_{\nu}^{\tau}	\\
{h_{\nu}^{e}s_{e}}	&	{h_{\nu}^{\mu}s_{\mu}}	&	{h_{\nu}^{\tau}s_{\tau}}	\\
0	&	0	&	0	\end{matrix}\right),
\end{equation}
and consequently the neutrino mass matrix is
\begin{equation}
\label{eq:Neutrino-mass-matrix_polar_parameters}
m_{\nu} = \frac{\mu_{\mathcal{N}} v_{3}^{2}}{{\left(h_{\mathcal{N}}^{1}\right)}^{2}v_{\chi}^{2}}
\left( \begin{matrix}
	\left({h_{\nu}^{e}}\right)^{2}	&h_{\nu}^{e}h_{\nu}^{\mu}c_{\nu}^{e\mu} 	&h_{\nu}^{e}h_{\nu}^{\tau}c_{\nu}^{e\tau}\\
	h_{\nu}^{e}h_{\nu}^{\mu}c_{\nu}^{e\mu}	&\left({h_{\nu}^{\mu}}\right)^{2}	&h_{\nu}^{\mu}h_{\nu}^{\tau}c_{\nu}^{\mu\tau}\\
	h_{\nu}^{e}h_{\nu}^{\tau}c_{\nu}^{e\tau}	&h_{\nu}^{\mu}h_{\nu}^{\tau}c_{\nu}^{\mu\tau}&\left({h_{\nu}^{\tau}}\right)^{2}
\end{matrix} \right),
\end{equation}
where $c_{\nu}^{\alpha\beta}=\cos(\theta_{\nu}^{\alpha}-\theta_{\nu}^{\beta})$. 
It is also possible to obtain the mass matrix by defining the following vectors in the neutrino Yukawa coupling space
\begin{equation}
\begin{split}
\mathbf{h}_{\nu}^{e} &= \left(h_{\nu}^{e}c_{\nu}^{e} , h_{\nu}^{e}s_{\nu}^{e}\right),	\\
\mathbf{h}_{\nu}^{\mu} &= \left(h_{\nu}^{\mu}c_{\nu}^{\mu} , h_{\nu}^{\mu}s_{\nu}^{\mu}\right),	\\
\mathbf{h}_{\nu}^{\tau} &= \left(h_{\nu}^{\tau}c_{\nu}^{\tau} , h_{\nu}^{\tau}s_{\nu}^{\tau}\right),
\end{split}
\end{equation}
in such a way that the mass matrix is obtained by dot-multiplying these vectors
\begin{equation}
m_{\nu} = \frac{\mu_{\mathcal{N}} v_{3}^{2}}{{\left(h_{\mathcal{N}}^{1}\right)}^{2}v_{\chi}^{2}}
\left( \begin{matrix}
	|\mathbf{h}_{\nu}^{e}|^{2}	&	\mathbf{h}_{\nu}^{e}\cdot\mathbf{h}_{\nu}^{\mu} 	&	\mathbf{h}_{\nu}^{e}\cdot\mathbf{h}_{\nu}^{\tau}	\\
	\mathbf{h}_{\nu}^{e}\cdot\mathbf{h}_{\nu}^{\mu}	&|\mathbf{h}_{\nu}^{\mu}|^{2}	&	\mathbf{h}_{\nu}^{\mu}\cdot\mathbf{h}_{\nu}^{\tau}	\\
	\mathbf{h}_{\nu}^{e}\cdot\mathbf{h}_{\nu}^{\tau}	&\mathbf{h}_{\nu}^{\mu}\cdot\mathbf{h}_{\nu}^{\tau}	&	|\mathbf{h}_{\nu}^{\tau}|^{2}
\end{matrix} \right).
\end{equation}

The new matrix can be diagonalized yielding the corresponding eigenvalues and eigenvectors, and also 
the mixing matrix and its angles using the definitions in \eqref{PMNS-mixing angles}. Moreover, in order to make consistent this model with neutrino oscillation data \cite{esteban2016updated,nufit}, the Yukawa parameters $(h_{\nu}^{e},\theta_{\nu}^{e})$, $(h_{\nu}^{\mu},\theta_{\nu}^{\mu})$, $(h_{\nu}^{\tau},\theta_{\nu}^{\tau})$ and $\theta_{12}^{E}$ should be fitted. Such a procedure is done with MonteCarlo method by generating one billion of trials in the parameter space and accepting points which match up the mass matrix to experimental data. It is worth mentioning that the other two rotation parameters $\theta_{13}^{E}$ and $\theta_{23}^{E}$ from eq. \eqref{eq:Electron-SM-Rotation-angles} were approximated to $m_{\tau}/m_{t}$. 

\begin{figure}[htbp]
\centering
\subfigure[${\left(h_{\mathcal{N}}^{1}\right)}^{2} = 0.01$.]{\includegraphics[width=45mm]{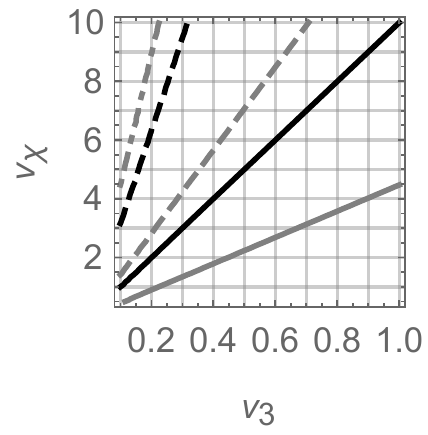}}
\subfigure[${\left(h_{\mathcal{N}}^{1}\right)}^{2} = 0.10$.]{\includegraphics[width=45mm]{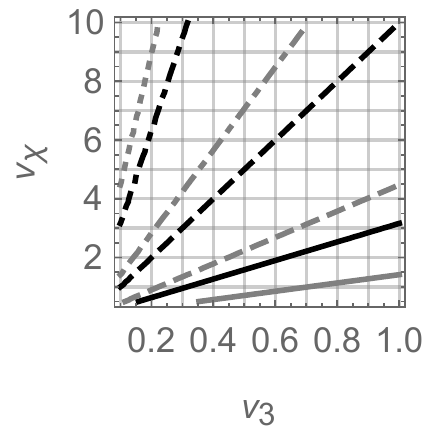}}
\subfigure[${\left(h_{\mathcal{N}}^{1}\right)}^{2} = 1.00$.]{\includegraphics[width=45mm]{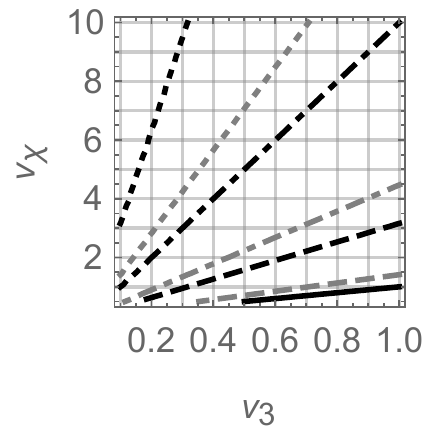}}
\caption{Contour plots of $v_{\chi}$ (TeV) vs. $v_{3}$ (GeV) from eq. \eqref{eq:Majorana-mass-NO-vChi-v2} for different values of ${\left(h_{\mathcal{N}}^{1}\right)}^{2}$ and $\mu_{\mathcal{N}}$. From below to above there are the corresponding contour plots for the following values of $\mu_{\mathcal{N}}$: 10 keV (gray, line), 50 keV (black, line), 100 keV (gray, dashed), 500 keV (black, dashed), 1 MeV (gray, dot-dashed), 5 MeV (black, dot-dashed), 10 MeV (gray, dotted) and 50 MeV (black, dotted).} 
\label{fig:external-factor-NO}
\end{figure}

On the other hand, the appropriate mass scale and mass ordering can be obtained by adjusting the outer factor of the mass matrix and the ratio $\rho$. For both NO and IO schemes, the Yukawa coupling constants can be set to
\begin{equation}
\label{eq:NO-IO-scale-constraint}
\begin{split}
{\left(h_{\mathcal{N}}^{1}\right)}^{2} &= 0.02,	\\
{\rho}^{2} &= 0.5,
\end{split}
\end{equation}
while the mass scale is set to
\begin{equation}
\begin{split}
v_{3} &= 0.5\mathrm{\,GeV},	\\
v_{\chi} &= 5\mathrm{\,TeV},	\\
\mu_{\mathcal{N}} &= 0.1\mathrm{\,MeV}.
\end{split}
\end{equation}
The above values fix the outer factor of the mass matrix \eqref{eq:Neutrino-mass-matrix} at $50\mathrm{\,meV}$, which yields to the correct squared-mass differences. Nevertheless, there exist other possible values for the parameters $\mu_{\mathcal{N}}$, $h_{N1\chi}$, $v_{\chi}$ and $v_{3}$ that gives the factor of $50$ meV. The only condition required to get the correct mass scale is
\begin{equation}
\label{eq:Majorana-mass-NO-vChi-v2}
\frac{\mu_{\mathcal{N}} v_{3}^{2}}{{\left(h_{\mathcal{N}}^{1}\right)}^{2}{v_{\chi}}^{2}} = 50\mathrm{\,meV}.
\end{equation}
By taking the above constraint and isolating $\mu_{\mathcal{N}}$, it is possible 
to obtain other solutions.
 These solutions are shown in the figure \ref{fig:external-factor-NO} where some contour plots on the $v_{\chi}$ vs. $v_{3}$ plane for different values of $\mu_{\mathcal{N}}$ from 10 keV to 50 MeV are shown. It is to note 
that due to the smallness of $v_{3}$, the Majorana mass scale $\mu_{\mathcal{N}}$ does not need to be small, contrary to other models \cite{catano2012neutrino,dias2012simple}. 
Specifically, values at MeV scale are consistent with the obsevable data. 

\begin{table*}
\centering
\begin{tabular}{c||ccc|cc}
$\theta_{12}^{E}$	&$h_{\nu}^{e}$	&$h_{\nu}^{\mu}$	&$h_{\nu}^{\tau}$	&$\theta_{\nu}^{\mu}$	&$\theta_{\nu}^{\tau}-\theta_{\nu}^{\mu}$\\\hline\hline
\multicolumn{6}{c}{Normal Ordering}\\ \hline\hline
$0^{\mathrm{o}}$	&	
\begin{tabular}{c}
$0.270\pm0.007$\\$0.271\pm0.007$\\
$0.274\pm0.007$\\$0.275\pm0.008$
\end{tabular}	&					
\begin{tabular}{c}					
$0.738\pm0.040$\\$0.741\pm0.041$\\
$0.737\pm0.043$\\$0.754\pm0.040$
\end{tabular}	&					
\begin{tabular}{c}						
$0.747\pm0.040$\\$0.745\pm0.041$\\
$0.745\pm0.043$\\$0.729\pm0.040$
\end{tabular}	&						
\begin{tabular}{c}						
$\pm(39.49\pm 2.99)$\\$\pm(140.24\pm2.93)$\\
$\pm(40.39\pm 2.80)$\\$\pm(78.17\pm 2.61)$
\end{tabular}	&					
\begin{tabular}{c}					
$\pm(38.79\pm 0.78)$\\$\mp(38.69\pm 0.84)$\\
$\mp(141.25\pm0.74)$\\$\mp(218.52\pm0.65)$
\end{tabular}\\\hline
$15^{\mathrm{o}}$	&
\begin{tabular}{c}
$0.294\pm0.008$\\
$0.362\pm0.014$\\$0.358\pm0.015$
\end{tabular}	&					
\begin{tabular}{c}					
$0.737\pm0.045$\\
$0.722\pm0.033$\\$0.720\pm0.036$
\end{tabular}	&					
\begin{tabular}{c}						
$0.738\pm0.043$\\
$0.725\pm0.041$\\$0.727\pm0.041$
\end{tabular}	&						
\begin{tabular}{c}						
$\pm(66.73\pm 1.02)$\\
$\pm(51.41\pm 2.81)$\\$\pm(50.75\pm 3.29)$
\end{tabular}	&					
\begin{tabular}{c}					
$\pm(33.32\pm 0.81)$\\
$\mp(43.99\pm 0.51)$\\$\mp(224.51\pm0.74)$
\end{tabular}\\\hline
$30^{\mathrm{o}}$	&	
\begin{tabular}{c}
$0.400\pm0.008$\\
$0.471\pm0.019$\\$0.402\pm0.010$
\end{tabular}	&					
\begin{tabular}{c}					
$0.689\pm0.035$\\
$0.625\pm0.021$\\$0.694\pm0.043$
\end{tabular}	&					
\begin{tabular}{c}						
$0.734\pm0.035$\\
$0.751\pm0.029$\\$0.729\pm0.045$
\end{tabular}	&						
\begin{tabular}{c}						
$\pm(46.38\pm 1.91)$\\
$\pm(42.39\pm 1.94)$\\$\pm(46.16\pm 2.21)$
\end{tabular}	&					
\begin{tabular}{c}					
$\pm(27.53\pm 1.12)$\\
$\pm(129.04\pm0.73)$\\$\mp(152.30\pm1.40)$
\end{tabular}\\\hline
$45^{\mathrm{o}}$	&	
\begin{tabular}{c}
$0.495\pm0.003$\\
$0.498\pm0.002$
\end{tabular}	&					
\begin{tabular}{c}					
$0.548\pm0.004$\\
$0.547\pm0.007$
\end{tabular}	&					
\begin{tabular}{c}						
$0.796\pm0.005$\\
$0.791\pm0.003$
\end{tabular}	&
\begin{tabular}{c}
$\pm(42.61\pm 0.82)$\\
$\pm(41.96\pm 0.78)$
\end{tabular}	&
\begin{tabular}{c}
$\pm(19.10\pm 0.75)$\\
$\pm(160.04\pm0.58)$
\end{tabular}\\\hline
\multicolumn{6}{c}{Inverted Ordering}\\ \hline\hline
$0^{\mathrm{o}}$	&	
\begin{tabular}{c}
$0.984\pm0.006$
\end{tabular}	&					
\begin{tabular}{c}					
$0.725\pm0.031$
\end{tabular}	&					
\begin{tabular}{c}						
$0.700\pm0.032$
\end{tabular}	&						
\begin{tabular}{c}						
$\pm(81.88\pm 0.84)$
\end{tabular}	&					
\begin{tabular}{c}					
$\mp(163.17\pm0.56)$
\end{tabular}\\\hline
$1^{\mathrm{o}}$	&	
\begin{tabular}{c}
$0.982\pm0.006$
\end{tabular}	&					
\begin{tabular}{c}					
$0.732\pm0.030$
\end{tabular}	&					
\begin{tabular}{c}						
$0.695\pm0.031$
\end{tabular}	&						
\begin{tabular}{c}						
$\pm(81.57\pm 0.74)$
\end{tabular}	&					
\begin{tabular}{c}					
$\pm(161.91\pm0.55)$
\end{tabular}\\\hline
$2^{\mathrm{o}}$	&	
\begin{tabular}{c}
$0.980\pm0.006$
\end{tabular}	&					
\begin{tabular}{c}					
$0.747\pm0.022$
\end{tabular}	&					
\begin{tabular}{c}						
$0.681\pm0.022$
\end{tabular}	&						
\begin{tabular}{c}						
$\pm(81.54\pm 0.51)$
\end{tabular}	&					
\begin{tabular}{c}
$\mp(160.77\pm0.50)$
\end{tabular}\\\hline
$3^{\mathrm{o}}$	&	
\begin{tabular}{c}
$0.978\pm0.006$
\end{tabular}	&					
\begin{tabular}{c}					
$0.759\pm0.014$
\end{tabular}	&					
\begin{tabular}{c}						
$0.671\pm0.013$
\end{tabular}	&						
\begin{tabular}{c}						
$\pm(81.51\pm 0.32)$
\end{tabular}	&					
\begin{tabular}{c}					
$\mp(159.56\pm0.46)$
\end{tabular}\\\hline
\end{tabular}
\caption{Parameter domains which reproduce neutrino data for NO and IO from \cite{esteban2016updated,nufit}. $\theta_{\nu}^{e}=0$.}
\label{tab:Neutrino-parameters}
\end{table*}

The table \ref{tab:Neutrino-parameters} shows the values of the parameters found with the MonteCarlo procedure consistent with the reported values in references \cite{esteban2016updated,nufit} for NO and IO. From the fact that the mass matrix is isotropic in the parameter space, $\theta_{\nu}^{e}$ was set to zero and any other solution with $\theta_{\nu}^{e}\neq 0$ is obtained by doing 3D rotations in the neutrino parameter space. The other angles are determined by $\theta_{\nu}^{\mu}$ and $\theta_{\nu}^{\tau}-\theta_{\nu}^{\mu}$. 

The NO scheme has several disconnected regions (there are twelve shown in table \ref{tab:Neutrino-parameters}) distributed in $3000$ solutions consistent with neutrino data which demonstrate the high consistency of the model with this scheme. Such values can be  replaced on \eqref{eq:Neutrino-mass-matrix_polar_parameters} in order to obtain the correct squared-mass differences and angles. However, the IO scheme has only four regions distributed in less than a thousand of solutions. It is due to the more restrictive constraints implied in this scheme. The masslessness of $\nu^{3}_{L}$, the values of the mixing angles and the quasi-degenerated masses of $\nu^{1}_{L}$ and $\nu^{2}_{L}$ restrict enormously the parameter regions, as it is shown in the table \ref{tab:Neutrino-parameters}. 

It is noteworthy the stringent constraint on $\theta_{12}^{E}$ in the IO scheme. The NO ones allows it to vary from $0^{\mathrm{o}}$ to $45^{\mathrm{o}}$, while the IO allows it to vary only from $0^{\mathrm{o}}$ to $3^{\mathrm{o}}$. The narrowing is observable in the $h_{\nu}^{\mu}$, $h_{\nu}^{\tau}$, $\theta_{\nu}^{\mu}$ and $\theta_{\nu}^{\tau}-\theta_{\nu}^{\mu}$ widths in the last four rows of the table \ref{tab:Neutrino-parameters}. On the contrary, there does not exist such stringent constraints in the NO scheme. 

\section{$h\rightarrow \tau \tau$ and $h\rightarrow \tau \mu$}
\label{sect:HLFV}
The Higgs Lepton Flavor Violation (HLFV) processes comprise some of the new hints in searching for new physics BSM. From them, the process $h\rightarrow \tau\mu$ suggests new physics since CMS 8 TeV had reported the branching ratio\cite{CMS_LFV}
\begin{equation}
\mathrm{BR}(h\rightarrow \tau\mu)=\left(0.84^{+0.39}_{-0.37}\right)\%.
\end{equation}
In comparison, ATLAS reported\cite{ATLAS_LFV}
\begin{equation}
\mathrm{BR}(h\rightarrow \tau\mu)=\left(0.53\pm0.51\right)\%,
\end{equation}
in consistence with CMS. Thus, the process $h\rightarrow \tau\mu$ may be given the first evidence on non-oscillatory LFV, together with the well-known neutrino oscillations. Now, according to the present model, the piece of the Lagrangian which predicts this process is
\begin{align}
\begin{split}
-\mathcal{L}_{E,\mathrm{SM},h} &= 
	\frac{h_{3e}^{ e  \mu }}{\sqrt{2}}\overline{e^{ e  }_{L}}h_{3}e^{\mu }_{R} + 
	\frac{h_{3e}^{\mu \mu }}{\sqrt{2}}\overline{e^{\mu }_{L}}h_{3}e^{\mu }_{R} \\ &+ 
	\frac{h_{2e}^{\tau  e }}{\sqrt{2}}\overline{e^{\tau}_{L}}h_{2}e^{ e  }_{R} + 
	\frac{h_{2e}^{\tau\tau}}{\sqrt{2}}\overline{e^{\tau}_{L}}h_{2}e^{\tau}_{R} + \mathrm{h.c.}
\end{split}
\end{align}
involving the interaction between charged leptons and Higgs doublets. Then, the replacement of the flavor states by the corresponding mass eigenstates with the rotation matrices
\begin{subequations}
\begin{align}
\mathbf{h}     &= R_{\mathrm{even}}^{\mathrm{T}}\mathbf{H}\\
\mathbf{E}_{L} &= \mathbb{V}^{E}_{L}\mathbf{e}_{L},	\\
\mathbf{E}_{R} &= \mathbb{V}^{E}_{R}\mathbf{e}_{R}.
\end{align}
\end{subequations}
is required in order to get the suited couplings. The rotation matrices can be expressed, at leading order, as 
\begin{subequations}
\begin{align}
R_{\mathrm{even}}^{\mathrm{T}}&=
\left(
\begin{array}{ccc}
 1 & -s_{12}^h & -s_{13}^h \\
 s_{12}^h & 1 & 0 \\
 s_{13}^h & 0 & 1 \\
\end{array}
\right)\\
\mathbb{V}^{E}_{L,\mathrm{SM}}&=
\left(
\begin{array}{ccc}
 \frac{1}{\sqrt{2}} & \frac{1}{\sqrt{2}} & s_{L,13}^E \\
 -\frac{1}{\sqrt{2}} & \frac{1}{\sqrt{2}} & s_{L,23}^E \\
 \frac{s_{L,23}^E-s_{L,13}^E}{\sqrt{2}} & -\frac{s_{L,23}^E+s_{L,13}^E}{\sqrt{2}} & 1 \\
\end{array}
\right),	\\
\mathbb{V}^{E}_{R}&=
\left(
\begin{array}{ccc}
 c_{R,12}^E & 1 & s_{R,13}^E \\
 -1 & c_{R,12}^E & s_{R,23}^E \\
 s_{R,23}^E & -s_{R,13}^E & 1 \\
\end{array}
\right).
\end{align}
\end{subequations}
After rotating the flavor basis into the mass eigenbasis, the Yukawa charged lepton Lagrangian can be expressed as
\begin{equation}
-\mathcal{L}_{E,\mathrm{SM},h}=y_{ij} \overline{e_{L}^{i}}e_{R}^{j}h + \mathrm{h.c.},
\end{equation}
where $y_{ii}$ ($y_{ij}$) are the conserving (violating) family lepton number coupling constants. 

\subsection{Conserving LN process $h\rightarrow \tau\tau$}
The conserving family lepton number process $h\rightarrow \tau\tau$ is predicted by the piece
\begin{equation}
-\mathcal{L}_{h\tau\tau} = y_{\tau\tau}\overline{\tau}\tau h
\end{equation}
with
\begin{align}
y_{\tau\tau} &= \frac{m_{\mu } s_{13}^h s_{R,13}^E}{\sqrt{2} \rho _3 v}+\frac{m_{\tau } s_{12}^h}{\rho _2 v}.
\end{align}
Thus, $s_{12}^h$ controls how much $h$ decays into $\tau\tau$. Additionally, the ratio
\begin{equation}
\frac{\sigma(h\rightarrow \tau\tau)}{\sigma(h\rightarrow \tau\tau)_{\mathrm{SM}}}=
\left\lbrace
\begin{split}
0.90 \pm 0.28			&\qquad \mathrm{CMS}	\\
1.43 \;\;^{+0.43}_{-0.37}	&\qquad \mathrm{ATLAS}
\end{split}
\right.
\end{equation}
comprises an important hint of BSM physics in flavor violation. In the present model, it turns out to be
\begin{equation}
\frac{\sigma(h\rightarrow \tau\tau)}{\sigma(h\rightarrow \tau\tau)_{\mathrm{SM}}}\approx
\left(\frac{s_{12}^{h}}{\rho_{2}}\right)^{2}.
\end{equation}
where $\rho_{2}\approx 0.05$ in order to obtain the masses of $b$ and $\tau$. The dependence on $s_{L,123}^{E}$, $s_{R,13}^{E}$, $s_{R,23}^{E}$ and $s_{13}^{h}$ are strongly suppressed by factors $m_{\mu}^{2}/m_{\tau}^{2}$. 
Therefore, the CMS and ATLAS limits yield the regions
\begin{equation}
s_{12}^{h} = \left\lbrace 
\begin{split}
\left(2.67\pm 0.85\right)\times 10^{-2} &\qquad \mathrm{CMS}	\\
\left(3.86\pm 0.06\right)\times 10^{-2} &\qquad \mathrm{ATLAS}
\end{split}
\right.
\end{equation}
constraining the available domains in the parameter space which is presented in the next paragraph in the light of the HLFV $h\rightarrow \tau\mu$ decay.

\subsection{HLFV $h\rightarrow \tau\mu$}
The process $h\rightarrow \tau\mu$ is predicted by the piece
\begin{equation}
-\mathcal{L}_{h\tau\mu} = y_{\mu\tau}\overline{\mu_{L}}\tau_{R}h + y_{\tau\mu}\overline{\tau_{L}}\mu_{R}h + \mathrm{h.c.}
\end{equation}
where the LFV couplings are
\begin{subequations}
\begin{align}
y_{\mu\tau} &= \frac{m_{\mu } s_{13}^h s_{R,23}^E}{\rho _3 v}-\frac{m_{\tau } s_{12}^h s_{L,123}^E}{\sqrt{2} \rho _2 v},\\
y_{\tau\mu} &= \frac{m_{\mu } s_{13}^h}{\sqrt{2} \rho _3 v}-\frac{m_{\tau } s_{12}^h s_{R,13}^E}{\rho _2 v},
\end{align}
\end{subequations}
and $s_{L,123}^E=s_{L,13}^E+s_{L,23}^E$. 
The BR from the Lagrangian is given by
\begin{equation}
\begin{split}
\mathrm{BR}(h\rightarrow \tau\mu) &= \frac{m_{h}}{8\pi \Gamma_{h}}\overline{y}_{\mu\tau}	
\approx 1200 \overline{y}_{\mu\tau},
\end{split}
\end{equation}
with the new parameter $\overline{y}_{\mu\tau}$ written as
\begin{equation}
\overline{y}_{\mu\tau} = \sqrt{y_{\mu\tau}^{2}+y_{\tau\mu} ^{2}}.
\end{equation}
According to \cite{herrero2016higgs}, the parameter $\overline{y}_{\mu\tau}$ lies in the region
\begin{equation}
0.002(0.001)<\overline{y}_{\mu\tau}<0.003(0.004)
\end{equation}
at 68\%(95\%) C.L. in such a way the CMS result might be explained. 

\begin{figure}[htbp]
\centering
\subfigure[]{\includegraphics[width=40mm]{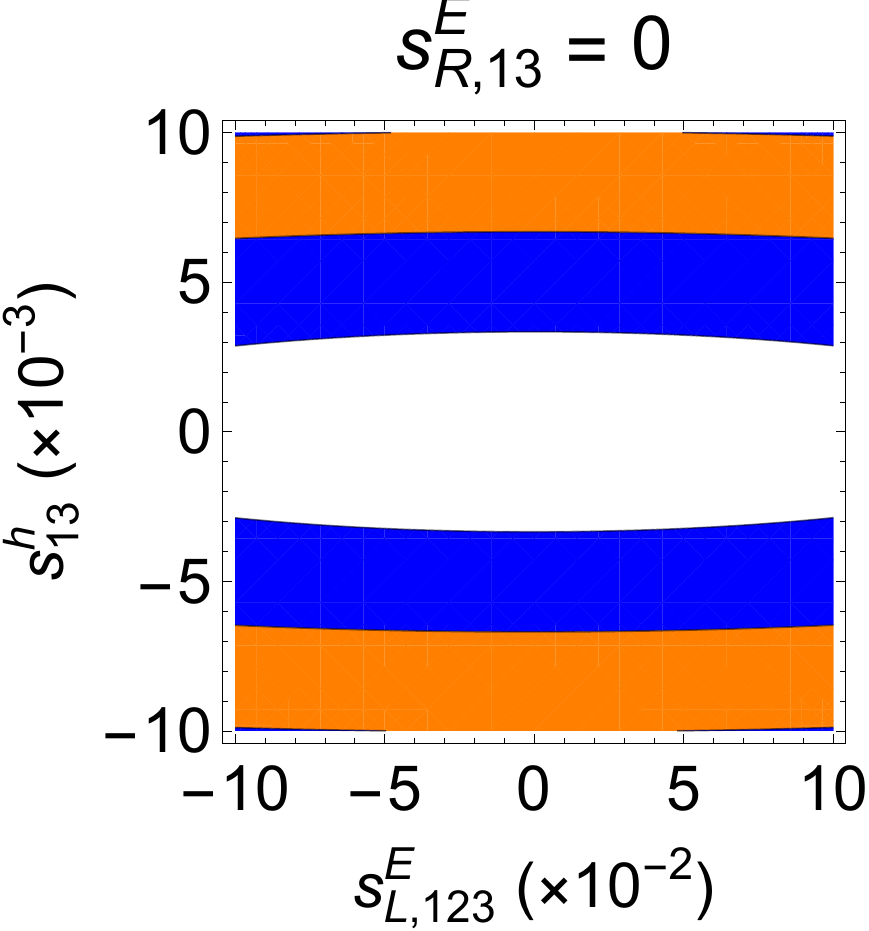}}
\subfigure[]{\includegraphics[width=40mm]{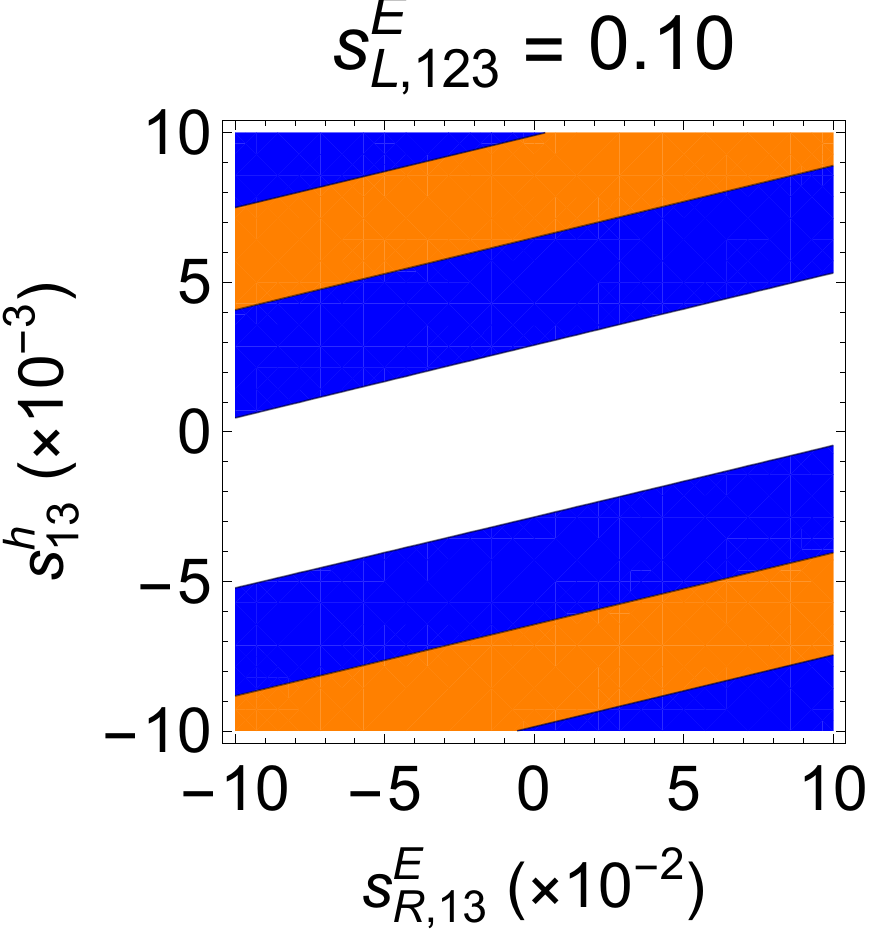}}
\caption{Contour plots of $\overline{y}_{\mu\tau}$ in the planes $s^{h}_{13}$ vs. $s_{L,123}^{E}$ and $s^{h}_{13}$ vs. $s_{R,13}^{E}$ with $s_{R,23}^{E}=0.01$ and $s_{12}^{h}=0.05$. The orange (blue) regions shows 68 \% (95 \%) C.L.} 
\label{fig:ContourSin13h}
\end{figure}

By fixing the value of $\overline{y}_{\mu\tau}$ at 68\% and 95\% C.L. some parameter spaces can be drafted to observe how much is consistent the model with the CMS report. First of all, $\overline{y}_{\mu\tau}$ does not depend on $s_{R,23}^{R}$ due to the $m_{\mu}/\rho_{3}v$ factor in the coefficient $y_{\mu\tau}$. Second, according to figure \ref{fig:ContourSin13h}a, the dependence on $s_{L,123}^{E}$ is slighter compared with $s_{13}^{h}$ and $s_{R,13}^{E}$ because of the constraint on $s_{12}^{h}\approx 0.05$. On the other hand, there exists a direct proportionality between $s_{13}^{h}$ and $s_{R,13}^{E}$ shown in figure \ref{fig:ContourSin13h}b in order to satisfy the CMS result. Moreover, it is observed how $s_{13}^{h}$ determines the actual value of $\overline{y}_{\mu\tau}$ in comparison with $s_{L,123}^{E}$ and $s_{13}^{h}$. Summarizing, $s_{13}^{h}$ should be larger than five in order to get the suited $\mathrm{BR}(h\rightarrow \tau\mu)$ in accordance with the CMS report. 


\section{Discussion and Conclusions}
\label{sect:Conclusion}
Some issues no explained by the SM, as the flavor problem, neutrino masses and mixing can be addressed employing the addition of abelian symmetries and the extension of the particle spectrum. The model shown here exhibits non-universal $\mathrm{U(1)}_{X}$ quantum numbers with $\mathbb{Z}_{2}$ parities which require extended scalar and fermion sectors in order to cancel chiral anomalies and to avoid massless charged fermions. The model implements three scalar doublets and one scalar singlet with an additional scalar field without VEV. The fermion sector includes three exotic quarks $\mathcal{T}$ and $\mathcal{J}^{1,2}$, five exotic leptons $\mathcal{E}^{1,2}$ and $\nu_{R}^{e,\mu,\tau}$, and three Majorana fermions $\mathcal{N}^{e,\mu,\tau}_{R}$ which allow different mass mechanisms in such a way that the fermion mass hierarchy is obtained naturally, as it was shown in section \ref{sect:Fermion-masses} in the equations \eqref{eq:Charged-Lepton-masses}, \eqref{eq:Up-Quarks-masses}, \eqref{eq:Down-Quarks-masses-Light} and \eqref{eq:Down-Quarks-masses-Heavy}. A summary about the fermion mass acquisition is presented in the table \ref{tab:Fermion-masses}.


\begin{table}[h]
\centering
\begin{tabular}{c cc cc}
	&	\multicolumn{2}{c}{Leptons}	&	\multicolumn{2}{c}{Quarks}\\ \hline\hline
Family	&	&	Mass	& 	&	Mass	\\ \hline\hline
1&$\nu_{L}^{1}$	&	$\dfrac{\mu_{\mathcal{N}} v_{3}^{2}}{{\left(h_{\mathcal{N}}^{1}\right)}^{2}v_{\chi}^{2}}h_{\nu1}^{2}$	&
$u$	&	$\dfrac{h_{u}^{2}-{h_{u}'}^{2}}{h_{t}}\dfrac{v_{3}}{\sqrt{2}} $\\
2&$\nu_{L}^{2}$	&	$\dfrac{\mu_{\mathcal{N}} v_{3}^{2}}{{\left(h_{\mathcal{N}}^{1}\right)}^{2}v_{\chi}^{2}}h_{\nu2}^{2}$	&
$c$	&	$\dfrac{h_{c}^{2}-{h_{c}'}^{2}}{h_{T}}\dfrac{v_{1}}{\sqrt{2}} $\\
3&$\nu_{L}^{3}$	&	$\dfrac{\mu_{\mathcal{N}} v_{3}^{2}}{{\left(h_{\mathcal{N}}^{1}\right)}^{2}v_{\chi}^{2}}h_{\nu3}^{2}$	&
$t$	&	$\dfrac{h_{t}v_{1}}{\sqrt{2}}$\\
Exot&\begin{tabular}{c}$N_{L}^{i}$ \end{tabular}	&	
\begin{tabular}{c}
$\dfrac{h_{\mathcal{N}}^{i}v_{\chi}}{\sqrt{2}}\mp\mu_{\mathcal{N}}$
\end{tabular}	&	
$T$	&	$\dfrac{h_{T}v_{\chi}}{\sqrt{2}}$\\	\hline
1&$e$	&	$\dfrac{h_{\ell}^{2}-{h_{\ell}'}^{2}}{h_{v1}}\dfrac{v_{3}}{\sqrt{2}} $	&
$d$	&	$\dfrac{\Sigma_{d}h_{d}^{2}}{h_{s}^{2}+{h_{s}'}^{2}}$\\
2&$\mu$	&	$\dfrac{h_{\ell}^{2}+{h_{\ell}'}^{2}}{h_{v1}}\dfrac{v_{3}}{\sqrt{2}} $	&
$s$	&	$\dfrac{h_{s}^{2}+{h_{s}'}^{2}}{h_{b}}\dfrac{v_{3}}{\sqrt{2}} $\\
3&$\tau$&	$\dfrac{h_{\tau}v_{2}}{\sqrt{2}}$	&
$b$	&	$\dfrac{h_{b}v_{2}}{\sqrt{2}}$\\
Exot&$E^{1}$&	$\dfrac{h_{E1} v_{\chi}}{\sqrt{2}}$	&
$J^{1}$	&	$\dfrac{h_{J1}v_{\chi}}{\sqrt{2}}$\\
Exot&$E^{2}$&	$\dfrac{h_{E2} v_{\chi}}{\sqrt{2}}$	&
$J^{2}$	&	$\dfrac{h_{J2}v_{\chi}}{\sqrt{2}}$\\	\hline
\end{tabular}
\caption{Summary of fermion masses showing their VEVs as well as the suppression mechanism if it is involved. The orders of magnitude of $v_{\chi}$, $v_{1}$, $v_{2}$, $v_{3}$ and $\mu_{\mathcal{N}}$ are units of TeV, hundreds of GeV, units of GeV, hundreds of MeV and units of MeV, respectively.}
\label{tab:Fermion-masses}
\end{table}

A vacuum hierarchy among the three Higgs doublets is obtained from the electroweak vacuum expectation value (VEV) $v=246$ GeV together with the third generation fermion masses, specially the $t$ quark mass at $m_{t}=173.21$ GeV. First, according to the SM, the electroweak bosons acquire mass through the VEVs of the three doublets, such that the effective electroweak VEV turns out to be
\begin{equation}
v^{2}=v_{1}^{2}+v_{2}^{2}+v_{3}^{2}=\left(246\mathrm{\,GeV}\right)^{2}.
\end{equation}
Second, the $t$ quark mass in the model is given by
\begin{equation}
m_{t}=\sqrt{(h_{1u}^{33})^{2}+(h_{1u}^{31})^{2}}\frac{v_{1}}{\sqrt{2}}=173.21\pm0.71\mathrm{\,GeV}.
\end{equation}
Additionally, if the Yukawa coupling constants are assumed at order one, we obtain that $v_{1}$ is close to the value of the electroweak VEV since $\sqrt{2}m_{t}\approx v_{1}$. Therefore, $v_{1}\approx 245.9$ GeV is the dominant contribution to the electroweak VEV, leaving a small gap to be filled by $v_{2}$ and $v_{3}$. Third, the $v_{2}$ and $v_{3}$ vacua are determined by the $b$-quark and $\tau$-lepton masses together with the muon and neutrino masses, respectively. Regarding to $v_{2}$, it is observed that $v_{2}\approx \sqrt{2}m_{b}\approx 6\mathrm{\,GeV}$, and since the $\tau/b$ mass ratio is of order 
\begin{equation}
\frac{m_{\tau}}{m_{b}} = 
\sqrt{\frac{\left(h_{2 e}^{\tau \tau}\right)^{2}}{(h_{2 d}^{3 3})^2+(h_{2 d}^{3 2})^2+(h_{2 d}^{3 1})^2}} \approx \frac{1}{2},
\end{equation}
the assumption to assign such a numerical value to $v_{2}$ is adequate. In addition, the fact that $v_{3}\approx \sqrt{2}m_{\mu}\approx 0.2\mathrm{\,GeV}$ with the neutrino mass scale factor obtained from the inverse see-saw mechanism
\begin{equation}
\frac{\mu_{\mathcal{N}}v_{3}^{2}}{v_{\chi}^{2}}=50\mathrm{\, meV},
\end{equation}
suggest that $v_{3}$ should be about $200\mathrm{\,MeV}$. Consequently, the vacuum hierarchy $v_{1} \ll v_{2} \ll v_{3}$ is consistent with the current phenomenological observations. 

On the other hand, the masses of the $u$-quark, $c$-quark and the electron appears as substractions between Yukawa coupling, that gives an additional suppresion of their masses. This feature is not accidental but come from the form of the see-saw formula (\ref{eq:see-saw-formula}):
\begin{equation*}
m^{\mathrm{sym}}_{F,\mathrm{SM}}\approx \mathcal{M}^{f}-\mathcal{M}^{f\mathcal{F}}\left(\mathcal{M}^{\mathcal{F}}  \right)^{-1}\mathcal{M}^{\mathcal{F}f},
\end{equation*}
and the vacuum hierarchy. For instance, the $c$ mass comes from the following sub-block in the up-like quark mass matrix
\begin{equation*}
\mathbb{M}_{c\mathcal{T}} \propto
\begin{pmatrix}
h_{1 u}^{2 2}v_{1}	&  \left| \right. & h_{1 \mathcal{T}}^{2}v_{1}	\\
\text{\textemdash}&\text{\textemdash}&\text{\textemdash}\\
g_{\chi u}^{2}v_{\chi}	&  \left| \right. & g_{\chi \mathcal{T}}v_{\chi}
\end{pmatrix}.
\end{equation*}
By taking into account the fact that $v_{\chi}\ll v_{1}$, 
the following eigenvalues are obtained:
\begin{equation*}
\begin{split}
m_{c}^{2}&=\frac{\left(h_{1 u}^{2 2}g_{\chi \mathcal{T}}-h_{1 \mathcal{T}}^{2}g_{\chi u}^{2}\right)^{2}}{(g_{\chi \mathcal{T}})^{2}+(g_{\chi u}^{2})^{2}}\frac{v_{1}^{2}}{2},	\\
m_{T}^{2}& = \left[(g_{\chi \mathcal{T}})^{2}+(g_{\chi u}^{2})^{2} \right]\frac{v_{\chi}^{2}}{2}+\frac{\left(h_{1 u}^{2 2}g_{\chi u}^{2}+h_{1 \mathcal{T}}^{2}g_{\chi \mathcal{T}}\right)^{2}}{(g_{\chi \mathcal{T}})^{2}+(g_{\chi u}^{2})^{2}}\frac{v_{1}^{2}}{2},
\end{split}
\end{equation*}
such that the exotic $\mathcal{T}$ quark suppress the $c$ quark mass. The same scenario appears in the $u$ quark mass, where the corresponding sub-block is
\begin{equation*}
\mathbb{M}_{ut} \propto
\begin{pmatrix}
h_{3 u}^{1 1}v_{3}	&\left| \right. & h_{3 u}^{1 3}v_{3}	\\
\text{\textemdash}&\text{\textemdash}&\text{\textemdash}\\
h_{1 u}^{3 1}v_{1}	&\left| \right. & h_{1 u}^{3 3}v_{1}
\end{pmatrix},
\end{equation*}
whose eigenvalues turn out to be (with the assumption $v_{3}/v_{1}\ll 1$):
\begin{equation*}
\begin{split}
m_{u}^{2}&=\frac{\left(h_{3 u}^{1 1}h_{1 u}^{3 3}-h_{3 u}^{1 3}h_{1 u}^{3 1}\right)^{2}}{(h_{1 u}^{3 3})^{2}+(h_{1 u}^{3 1})^{2}}\frac{v_{3}^{2}}{2},	\\
m_{t}^{2}&=\left[(h_{1 u}^{3 3})^{2}+(h_{1 u}^{3 1})^{2} \right]\frac{v_{1}^{2}}{2}+\frac{\left(h_{3 u}^{1 1}h_{1 u}^{3 3}+h_{3 u}^{1 3}h_{1 u}^{3 1}\right)^{2}}{(h_{1 u}^{3 3})^{2}+(h_{1 u}^{3 1})^{2}}\frac{v_{3}^{2}}{2},\\
\end{split}
\end{equation*}
and it is observed again how a heavier mass, in this case the $t$ quark mass, suppress the mass of the light $u$ quark. Finally, the $e$ and $\mu$ leptons shows a similar behavior, but the suppression is between $v_{3}$ in the second column with the $v_{1}$ in the fourth column of the charged lepton mass matrix
\begin{equation}
\mathbb{M}_{e\mu} \propto
\begin{pmatrix}
h_{3 e}^{e \mu}v_{3}	&\left| \right. & h_{1 \mathcal{E}}^{e 1}v_{1}	\\
\text{\textemdash}&\text{\textemdash}&\text{\textemdash}\\
h_{3 e}^{\mu \mu}v_{3}	&\left| \right. & h_{1 \mathcal{E}}^{\mu 1}v_{1}
\end{pmatrix},
\end{equation}
such that the masses of the lightest charged leptons are
\begin{equation}
\begin{split}
m_{e}^{2} &= \frac{\left(h_{3 e}^{e \mu}h_{1 \mathcal{E}}^{\mu 1}-h_{3 e}^{\mu \mu}h_{1 \mathcal{E}}^{e 1}\right)^2}{(h_{1 \mathcal{E}}^{e 1})^2+(h_{1 \mathcal{E}}^{\mu 1})^2}\frac{v_{3}^{2}}{2},		\\
m_{\mu}^{2} &= \frac{\left(h_{3 e}^{e \mu}h_{1 \mathcal{E}}^{e 1}+h_{3 e}^{\mu \mu}h_{1 \mathcal{E}}^{\mu 1}\right)^2}{(h_{1 \mathcal{E}}^{e 1})^2+(h_{1 \mathcal{E}}^{\mu 1})^2}\frac{v_{3}^{2}}{2}.	\\
\end{split}
\end{equation}
These suppression mechanisms are induced by the vacuum hierarchy together with the zero texture matrices obtained from the non-universal $\mathrm{U(1)}_{X}$ interaction and the $\mathbb{Z}_{2}$ parity. Consequently, the subtraction of Yukawa coupling constants is a natural consequence of the diagonalization of the mass matrices by taking into account the vacuum hierarchy outlined in the previous paragraphs. 

Furthermore, the model is not only consistent with the fermion mass hierarchy of the SM, but also it is consistent with some phenomenological reports in the lepton sector. Regarding to the consistency of the model with current neutrino oscillation data, this scheme is consistent with both mass orderings, NO and IO, but the former is preferred because of the large abundance of solutions in comparison with the latter. The PMNS angles and the squared-mass differences are satisfied without needing fine-tunnings due to the fact that the parameters $h^{e}_{\nu}$, $h^{\mu}_{\nu}$ and $h^{\tau}_{\nu}$ vary from zero to one, and similarly the angles $\theta^{\mu}_{\nu}$ and $\theta^{\tau}_{\nu}$ span $\pm 180^{\mathrm{o}}$. Moreover, the suited neutrino mass scale is fitted by the Majorana mass $\mu_{\mathcal{N}}$, the Yukawa coupling $h_{\mathcal{N}}^{1}$ and the vacua $v_{3}$ and $v_{\chi}$, so the model has a large set of solutions in order to be consistent with neutrino oscillation data. Additionally, the model is adecquate to understand the CMS report about the $h\rightarrow \tau\mu$ branching ratio. The mixing angles of the CP-even scalars, the left- and right-handed charged leptons, together with the vacuum hierarchy yield definite regions of consistency in the mixing angle space, where the most important relation is between $s_{13}^{h}$ and $s_{R,13}^{E}$. 

Regarding the exotic neutral sector $N$ and $\widetilde{N}$, the equations \eqref{eq:R-Neutrino-mass-matrix} and \eqref{eq:NO-IO-scale-constraint} allow us to set the pseudo-Dirac neutrinos masses from 100 GeV to 700 GeV by setting $\mu_{\mathcal{N}}$ from 4 keV to 0.1 GeV such that $N$ and $\widetilde{N}$ can be observed at current energy scales at particle colliders. Moreover, since $h_{\mathcal{N}}^{3}$ in eq. \ref{eq:R-Neutrino-mass-matrix} does not matter in setting the correct squared-mass differences of light neutrinos, $N^{3}$ and $\widetilde{N}^{3}$ masses are not constrained. 

The present model shows how the introduction of new non-universal quantum numbers and an extended scalar sector present a fertile scenario where some issues, which the SM cannot explained such as the fermion mass hierarchy, HLFV processes or, last but not least, the evidence of the massive nature of neutrinos in their oscillations, can be understood with the introduction of the least number of new particles and symmetries. 

\section*{Acknowledgments}
This work was supported by \textit{El Patrimonio Aut\'onomo Fondo Nacional de Financiamiento para la Ciencia, la Tecnolog\'{i}a y la 	Innovaci\'on Francisco Jos\'e de Caldas} programme of COLCIENCIAS in Colombia. RM thanks to professor German Valencia for the kindly hospitality at Monash University and his useful comments.
\vspace{0.2cm}

\bibliography{3HD+1HSnonuniversal}
\bibliographystyle{ieeetr}

\end{document}